\documentclass{article}
\usepackage[margin=3cm]{geometry}

\usepackage{authblk}

\usepackage{amsmath,amssymb,amsfonts}
\usepackage{mathtools} 
\usepackage{physics}
\usepackage{makecell}
\setcellgapes{1ex}
\usepackage{cases}
\usepackage{booktabs} 
\usepackage{multirow}

\usepackage{graphicx,color}
\usepackage{epstopdf}
\usepackage{caption}
\usepackage{subcaption}
\usepackage[nocompress]{cite}

\usepackage[shortlabels]{enumitem}

\usepackage{caption}
\usepackage{subcaption}

\usepackage{float}

\usepackage{todonotes}

\newcommand{\ourG}{\mathbf{G}}
\newcommand{\ourB}{\mathbf{B}}

\newcommand{\LL}{\mathcal{L}}
\renewcommand{\vec}{\mathbf}
\newcommand{\dalam}{\raisebox{-.75px}{$\Box$}}

\numberwithin{equation}{section}

\usepackage{hyperref}
\hypersetup{colorlinks=true,allcolors=blue,urlcolor=cyan}

\title{Cosmological dynamical systems of non-minimally coupled fluids and scalar fields}

\author[1]{H. A. Ashi\footnote{Email: haashi@kau.edu.sa}}
\author[2,3]{Christian G. B\"ohmer\footnote{Email: c.boehmer@ucl.ac.uk}}
\author[2]{Antonio d'Alfonso del Sordo\footnote{Email: a.dalfonsodelsordo@ucl.ac.uk}}
\author[2]{\ Erik Jensko\footnote{Email: erik.jensko@ucl.ac.uk (\textbf{corresponding author})}}

\affil[1]{Department of Mathematics, \authorcr King Abdulaziz University, Jeddah, Saudi Arabia\medskip}
\affil[2]{Department of Mathematics, University College London, \authorcr Gower Street, London WC1E 6BT, UK\medskip}
\affil[3]{Astrophysics Research Centre, School of Mathematics, \authorcr Statistics and Computer Science, University of KwaZulu-Natal, \authorcr Private Bag X54001, Durban 4000, South Africa\medskip}

\date{\today} 

\pagecolor{white}

\begin{document}

\renewcommand{\arraystretch}{1.2} 
\setlength{\tabcolsep}{1ex} 
\setlength{\extrarowheight}{1ex} 

\maketitle

\begin{abstract}
We study the cosmological dynamics of non-minimally coupled matter models using the Brown's variational approach to relativistic fluids in General Relativity. After decomposing the Ricci scalar into a bulk and a boundary term, we construct new models by coupling the bulk term to the fluid variables and an external scalar field. Using dynamical systems techniques, we study models of this type and find that they can give rise to both early-time inflationary behaviour and late-time accelerated expansion. Moreover, these models also contain very interesting features that are rarely seen in this context. For example, we find dark energy models which exhibit phantom crossing in the recent past. Other possibilities include models that give a viable past evolution but terminate in a matter-dominated universe. The dynamical systems themselves display an array of mathematically interesting phenomena, including spirals, centres, and non-trivial bifurcations depending on the chosen parameter values.
\end{abstract}

\clearpage 
\tableofcontents
\clearpage

\section{Introduction}

The nature of dark energy and dark matter remains one of the most pressing topics in modern cosmology. The $\Lambda$CDM model successfully describes the large-scale structure and the accelerated expansion of the universe~\cite{SupernovaSearchTeam:1998fmf,SupernovaCosmologyProject:1998vns,LIGOScientific:2016aoc,Planck:2018vyg,Will:2018bme,Ishak:2018his,Zlatev:1998tr,Sadjadi:2006qp}. However, the model struggles to explain the observed value of the cosmological constant, evidenced by the infamous cosmological constant problem \cite{Weinberg:1988cp}. 
Additionally, the comparable energy densities of dark matter and dark energy today present the coincidence problem, suggesting a dynamical explanation for dark energy and potential interactions between these components.

Further challenges to the $\Lambda$CDM model come from the ongoing observational tensions in cosmology~\cite{Abdalla:2022yfr,CosmoVerseNetwork:2025alb}, the most prominent being the Hubble tension; this discrepancy between the early-time inferred value of $H_0$ and the late-time direct measurements has been shown to be especially difficult to reconcile in the standard $\Lambda$CDM framework~\cite{Verde:2019ivm,DiValentino:2020zio}. To add to this, the recent surveys of the Dark Energy Spectroscopic Instrument (DESI) show tensions in the $\Lambda$CDM model when comparing baryon acoustic oscillations (BAO), Cosmic Microwave Background (CMB) and supernovae data~\cite{DESI:2025zpo,DESI:2025zgx}. In fact, these results point towards dynamical dark energy with a phantom-like equation of state in the recent past---a result that appears to hold across many different  parametrisations~\cite{Pan:2025psn,Li:2025ops}.

To address these issues, alternative models have been proposed, including dynamical dark energy (e.g.~quintessence), modified gravity theories, and interacting dark sector scenarios~\cite{Copeland:2006wr,Oks:2021hef,Wang:2016lxa}. Among these, scalar field models coupled to matter have been particularly fruitful~\cite{Tsujikawa:2013fta}, offering mechanisms for cosmic acceleration and alleviating fine-tuning problems. Early-time accelerated expansion models can also be constructed using scalar fields~\cite{Tamanini:2014mpa,Bohmer:2016ome,Urena-Lopez:2020npg}.
However, many existing coupled models rely on \emph{ad hoc} phenomenological couplings rather than deriving interactions from fundamental principles. To combat these limitations, variational approaches such as modifications of the Schutz and Sorkin action~\cite{Schutz:1977df} have been proposed to formulate interacting cosmological models, see~\cite{Pourtsidou:2013nha,Skordis:2015yra,Amendola:2020ldb,Kase:2020hst,Kase:2019veo,BeltranJimenez:2020qdu,BeltranJimenez:2021wbq,DeFelice:2016yws,Aoki:2025bmj}. A different variational approach for coupling scalar fields to matter was put forward in~\cite{Boehmer:2015kta,Boehmer:2015sha}, based on the Brown fluid Lagrangian~\cite{Brown:1992kc}. Using this approach, new couplings related to the boundary terms introduced in~\cite{Boehmer:2021aji,Boehmer:2023fyl} were studied using the modified Brown Lagrangian formalism~\cite{Boehmer:2024rqk}. These couplings were shown to lead to a rich structure and a variety of interesting dynamics in cosmological contexts.

In this work, we continue to investigate models which contain non-minimal couplings between matter and geometry. We consider an interaction term that depends on the matter particle number density, an external scalar field, and a geometrical term labelled as $\ourG$. This term $\ourG$ appears in the Einstein action and yields the standard Einstein field equations, differing from the Ricci scalar by a boundary term~\cite{Boehmer:2021aji}; however, this term depends only on first derivatives of the metric and is not a coordinate scalar. Consequently, when non-minimally coupled to matter, the resulting models deviate significantly from those constructed from the Ricci scalar. A key difference to the latter is the apparent breaking of diffeomorphism invariance, which can be interpreted as higher-order effective corrections to standard General Relativity \cite{Donoghue:2009mn}. Alternatively, the non-minimal couplings can be viewed in a covariant manner as simply introducing new degrees of freedom; this is made explicit when taking the viewpoint of symmetric teleparallel gravity, where the new degrees of freedom are associated with Stueuckelberg fields related to nonmetricity. In this approach, the geometric term $\ourG$ is simply the nonmetricity scalar $Q$ fixed to the coincident gauge \cite{Jensko:2024bee,BeltranJimenez:2022azb}. 
It follows that the results of this paper can equally be applied to the symmetric teleparallel theories of gravity, broadening the scope of this work.

We utilise dynamical systems techniques~\cite{Bahamonde:2017ize}, which has been successfully applied to various interacting dark sector models~\cite{Ashmita:2024ueh,Kritpetch:2024rgi}. We analyse the background cosmology of these models and identify the fixed points of the system. Many phenomenologically interesting points are found, such as corresponding to accelerated expansion, scaling solutions, or matter-dominated epochs. The resulting models contain a rich phase space structure, allowing for both early-time inflation and late-time acceleration. Linear stability analyses are then performed to confirm the asymptotic behaviour of each model, after which we study the evolution of physical parameters for a range of initial conditions.

The non-minimally coupled framework we construct is very general, allowing for a wide variety of different types of couplings. Even the simplest coupling gives rise to an interesting array of dynamics. Quite remarkably, as well as the standard late-time $\Lambda$CDM scenario (characterised by a matter-dominated period followed by de Sitter expansion), the models can give rise to a viable transient phantom dark energy phase; this type of behaviour is especially interesting in light of the recent DESI observational results~\cite{DESI:2025zpo,DESI:2025zgx}.
Other scenarios follow a $\Lambda$CDM trajectory before displaying tracking behaviour, ending in a matter-dominated universe.

By formulating dark sector interactions though a variational principles and geometric boundary terms, this work opens new avenues for exploring fundamental cosmology beyond $\Lambda$CDM, with potential connections to modified gravity and phenomenological models displaying similar features.

The paper is organised as follows. In Section~\ref{sec:lagrangian}, we present the Lagrangian formulation of our framework. In Section~\ref{sec:cosmology}, we focus on the cosmological field equations and introduce the cosmological variables which will be used for our analysis. Section~\ref{sec:phase} contains our analysis of the dynamical systems for a model which depends on two new parameters. Finally, in Section~\ref{sec:disc}, we present a discussion of our results and future investigations. 

\paragraph{Notation and conventions.}
Unless otherwise specified, we employ standard relativistic notation throughout.  The signature of the metric tensor $g_{\mu\nu}$ is $(-,+,+,+)$, $g$ denotes the determinant of the metric, and Greek indices are space-time indices taking values $(0,1,2,3)$. The gravitational coupling constant is $\kappa=8\pi G/c^4$, where $c$ is the speed of light and $G$ the Newton's gravitational constant. Throughout, we use natural units with $c=1$ together with $\kappa=M_{\mathrm{Pl}}^{-2}=8\pi G$. A dot will denote differentiation with respect to cosmological time. A prime denotes the derivative with respect to the argument. Where convenient, the comma notation for partial derivatives is used $\phi_{,\mu}=\partial_\mu \phi$. 

\section{Lagrangian formulation}
\label{sec:lagrangian}

We begin by briefly reviewing the variational approach to relativistic fluids in General Relativity, followed by a short introduction to the key quantities which appear in the Einstein action. Once this general framework is introduced, it becomes straightforward to state an interaction term. We then comment on the relation between our chosen interaction Lagrangian and similar constructions with boundary terms.

\subsection{Fluid and scalar field action}

The Lagrangian formulation for perfect fluids introduced by Brown~\cite{Brown:1992kc} is derived from the Lagrangian density
\begin{equation}
    \mathcal{L}_{\textrm{fluid}} = -\sqrt{-g} \rho(n,s) + J^\mu \left(\varphi_{,\mu} + s \theta_{,\mu} + \beta_{A}\alpha^{A}_{, \mu} \right)\, ,
\end{equation}
where 
\begin{itemize}
    \item $n$ is the particle number density
    \item $s$ is entropy density per particle
    \item $\rho(n,s)$ is the energy density of the matter fluid 
    \item  $J^\mu$ is the particle-number flux density, which is related to $n$ by
    \begin{equation}
        J^\mu=\sqrt{-g}\,n\,U^\mu, \qquad 
        |J|=\sqrt{-J_{\mu} J^\mu}, \qquad 
        n=|J|/\sqrt{-g}, 
        \label{defnofJandn}
    \end{equation}
    and $U^\mu$ is the fluid's $4$-velocity satisfying $U_\mu U^\mu=-1$
    \item  $\varphi$, $\theta$, and $\beta_A$ are Lagrange multipliers and $\alpha^A(x^\mu)$ are the Lagrangian coordinates of the fluid which are functions of the spacetime coordinates.
\end{itemize}
The Lagrange multipliers $\varphi$ and $\theta$ serve to enforce the conservation of particle number and entropy flux along the fluid flow lines, while $\beta_A$ fixes the fluid 4-velocity to be directed along the flow lines.

In addition, we define the thermodynamic quantities 
\begin{align} 
    \label{pressure}
    \textrm{pressure:} \qquad p &= 
    n \frac{\partial \rho}{\partial n} - \rho \,,\\
    \textrm{temperature:} \qquad T &= 
    \frac{1}{n} \frac{\partial \rho}{\partial s} \,, \\
    \textrm{chemical potential:} \qquad \mu &= 
    \frac{\rho + p}{n} \,,
\end{align}
which then agree with the first law of thermodynamics 
\begin{equation}
    \dd \rho = \mu\, \dd n + T\, \dd s \,.
\end{equation}
The Lagrange multipliers $\varphi$ and $\theta$ can also be given thermodynamic interpretations, relating to the thermasy and chemical momentum of the fluid~\cite{VANDANTZIG1939673,Brown:1992kc}. Theories with modification of these Lagrange multiplier terms have been considered in~\cite{Iosifidis:2024ksa}, leading to models with non-conservative and non-adiabatic effects. 

In addition to the fluid matter action, let us also introduce a canonical scalar field with Lagrangian density
\begin{equation}
    \mathcal{L}_\phi = -\sqrt{-g}\Bigl(\tfrac{1}{2}\nabla_\mu\phi\nabla^\mu\phi+V(\phi)\Bigr) \,,
\end{equation}
where $V(\phi)$ is the scalar field's potential. Just like in the usual quintessence scenario, the scalar field will play the role of dark energy and provide a mechanism for the accelerated expansion of the universe. However, when non-minimal couplings are introduced in the next section, the interpretation of the scalar field will be more subtle, with it entering into the definition of the effective energy-momentum tensor. It is also worth noting that interacting models with modified scalar field kinetic and potential terms have been considered in the literature~\cite{Pourtsidou:2013nha,Skordis:2015yra}. In this work, we leave the canonical scalar field Lagrangian unmodified. 

\subsection{Gravitational action and interaction term}

Inspired by~\cite{Boehmer:2024rqk}, we consider an interesting new type of coupling to matter involving non-covariant terms built from the metric tensor and its derivatives. These non-covariant terms are derived from deconstructing the Ricci scalar into a bulk and boundary part
\begin{equation}
    R = \ourG + \ourB  \,,
\end{equation}
where the first-order term $\ourG$ and the boundary term $\ourB$ are given by
\begin{align}
    \ourG &:= g^{\mu\nu} (\Gamma^\lambda_{\mu\sigma}\Gamma^\sigma_{\lambda\nu}-\Gamma^\sigma_{\mu\nu}\Gamma^\lambda_{\lambda\sigma}) \,, 
    \label{bulk}\\
    \ourB &:= \frac{1}{\sqrt{-g}} \partial_{\nu}(\sqrt{-g} B^{\nu}) \,, \label{boundary}
\end{align}
where $B^{\nu} = 2 g^{\mu [\lambda} \Gamma^{\nu]}_{\mu \lambda}$ is the boundary pseudovector defined in terms of the Levi-Civita connection~\cite{Boehmer:2021aji}. Recall that square brackets around indices stand for skew-symmetrisation. The bulk part is sometimes known as the \textit{Einstein action} as its variations lead to the Einstein field equations. The boundary term $\ourB$ is a total derivative and hence does not play a dynamical role.  It should then be clear that while $\ourG$ leads to covariant field equations when uncoupled, this does not hold for non-linear functions of $\ourG$ or when non-minimal couplings are considered.

Modifications of gravity based on this decomposition have been considered in~\cite{Boehmer:2021aji,Boehmer:2023fyl,Jensko:2023lmn}, where non-linear functions that break covariance are examined. These theories give rise to extended models beyond GR, which also differ from other popular modifications such as $f(R)$ gravity.
Further discussion on the consistency of these diffeomorphism-breaking approaches and the role of coordinates in these models can be found in~\cite{Jensko:2024bee}. Most relevant to our work are the extensions with non-minimal couplings to the boundary term $\ourB$~\cite{Boehmer:2024rqk}. There, the authors studied models with an interaction Lagrangian with derivative couplings of the form $f(n,s,\phi)B^{\mu}\partial_{\mu}$. These derivative couplings are also closely related to algebraic models of the form $f(n,s,\phi,\ourB)$, which we discuss in Section~\ref{algeb}.
Here, we will instead couple the scalar field $\phi$ and fluid matter variables to the bulk term $\ourG$, marking a considerable departure from~\cite{Boehmer:2024rqk}. This will be reflected in the subsequent dynamical systems analysis which differs from the previous work significantly. 

The total action is given by
\begin{align}
    S_{\mathrm{tot}}= \int \left( \LL_{\mathrm{EH}}+\LL_{\mathrm{fluid}}+\LL_{\phi}+\LL_{\mathrm{int}}\right) \dd^4 x \,, 
    \label{totalaction}
\end{align}
where $\mathcal{L}_{\textrm{EH}}$ is the Einstein-Hilbert Lagrangian density and the interaction term $\LL_{\mathrm{int}}$ is a function of $n$, $s$, $\phi$ and $\ourG$. In this work, we will study algebraic couplings of the form
\begin{equation}
    \LL_{\mathrm{int}} = - \sqrt{-g}f(n,s,\phi,\ourG) \, ,
\end{equation}
such that the equations of motion will remain at most second-order in metric derivatives. We use the phrase `algebraic coupling' to distinguish these models from couplings which contain derivatives. An example of the latter would be a coupling of the form $f(n,s,\ourG)J^\mu\partial_\mu \phi$, similar to the models studied in~\cite{Boehmer:2015sha}, where couplings of this type were considered without additional geometrical couplings. We also point out non-minimally coupled models using the Brown fluid variables $n$, $s$ have not been studied in the analogous $f(T)$ and $f(Q)$ gravity frameworks, marking the novelty of our model.

The action~(\ref{totalaction}) depends on the variables $g_{\mu \nu}$, $s$, $J^{\mu}$, $\varphi$, $\theta$, $\alpha^{A}$, $\beta_{A}$, $\phi$, which must all be varied independently. Before calculating the variations, let us comment on the consistency of such a theory with non-covariant terms, which is intimately related to conservation laws.

In interacting models with nominally coupled terms, the covariant conservation of the total energy-momentum $\nabla^{\mu} T_{\mu \nu}^{\textrm{total}}=0$ can be seen as a consequence of the diffeomorphism invariance of the matter action $\delta_{\xi} S_{\textrm{matter}} =0$~\cite{Koivisto:2005yk}. On the other hand, including diffeomorphism-breaking interaction terms $\ourG$ or $\ourB$ implies that infinitesimal variations of the action $\delta_{\xi} S_{\textrm{matter}}$ do not vanish identically. The form of the interaction term $f(n,s,\phi,\ourG)$ in~(\ref{totalaction}) must then be constrained in order to preserve the total covariant conservation of matter
\begin{align} 
    \label{noncons}
    \nabla^{\mu} T_{\mu \nu}^{\textrm{total}}= \nabla^{\mu} \big( T^{(\rm fluid)}_{\mu\nu} + T_{\mu\nu}^{(\phi)} + T_{\mu\nu}^{(\mathrm{int})} \big) = 0 \,.
\end{align}
The above conservation law also follows directly from the metric field equations and the contracted Bianchi identity $\nabla^{\mu}G_{\mu \nu}=0$. In Appendix \ref{append_diff}, we show that the conservation equation is indeed satisfied for all choices of interaction function $f$ on FLRW backgrounds in Cartesian coordinates. In fact, the analysis in Appendix \ref{append_diff} is fully general and applies to all models constructed from either $\ourG$ or $\ourB$. Hence, this explains why past works constructed from these objects~\cite{Boehmer:2021aji,Boehmer:2023fyl,Boehmer:2024rqk} bypass any potential issues that may arise from breaking diffeomorphism invariance. Similar consistency conditions were found in the diffeomorphism-breaking models studied in~\cite{Bello-Morales:2023btf,Anber:2009qp}. We are therefore free to choose any form of the function $f$ in the subsequent analysis.

\subsection{Equations of motion}

Beginning with the Lagrange multipliers in the matter sector, we find 
\begin{align}
    \delta\varphi:&\quad J^\mu_{,\mu}=0 \, ,
    \label{varwrtvarphi}\\
    \delta\theta:&\quad \left(sJ^\mu\right)_{,\mu}=0 \, ,
    \label{varwrttheta}\\
    \delta\alpha^A:&\quad \left(\beta_A J^\mu\right)_{,\mu}=0 \, ,
    \label{varwrtalphaA}\\
   \delta\beta^A:&\quad \alpha^A_{,\mu} J^\mu=0 \, .
   \label{varwrtbetaA}
\end{align}
These equations are independent of the gravitational actions and are also independent of the interaction term. The first two equations enforce the conservation of particle number and entropy along the fluid flow respectively
\begin{equation}
    \nabla_{\mu}(n U^{\mu}) =0 \, , \quad \nabla_{\mu}(n s  U^{\mu}) =0 \, , \label{nscons}
\end{equation}
from which we have $U^{\mu} \partial_{\mu} s = 0$. 
The variations with respect to the entropy $s$ and flux $J^{\mu}$ are both modified by the interaction term
\begin{align}
    \delta s:&\quad \frac{\partial\rho}{\partial s} -nU^\mu\theta_{,\mu} +\frac{\partial f}{\partial s} \,, \\ 
    \delta J^\mu:&\quad \varphi_{,\mu}+s\theta_{,\mu}+\beta_A\alpha^A_{,\mu}+\frac{\partial \rho}{\partial n}U_{\mu}+\frac{\partial f}{\partial n}U_{\mu} =0 \,.
\end{align}
Variations with respect to the scalar field $\phi$ yield a modified Klein Gordon equation
\begin{equation} 
    \label{scalar_EoM}
     \delta\phi:\quad \dalam\phi-V'(\phi)- \frac{\partial f}{\partial \phi}=0  \, ,
\end{equation}
where $\dalam \phi := \nabla^{\mu} \nabla_{\mu} \phi$.
Finally, variations with respect to the metric yield the modified Einstein field equations 
\begin{equation} \label{met_EoM}
  \delta g^{\mu\nu}:\quad G_{\mu\nu}=
  \kappa \big( T^{(\rm fluid)}_{\mu\nu} + T_{\mu\nu}^{(\phi)} + T_{\mu\nu}^{(\mathrm{int})} \big) \, ,
\end{equation}
where $G_{\mu\nu}$ is the Einstein tensor. The energy-momentum tensor for the fluid, scalar field and interaction terms are  
\begin{align}
    T^{(\rm fluid)}_{\mu\nu} &= 
    (\rho+p)U_{\mu}U_{\nu} + p g_{\mu\nu} \, , \\
    T_{\mu\nu}^{(\phi)} &= 
    \partial_\mu \phi\partial_\nu \phi -
    \frac{1}{2} g_{\mu\nu}\partial_\lambda\phi\partial^\lambda\phi -g_{\mu\nu} V(\phi) \, , \\
    T_{\mu\nu}^{(\mathrm{int})} &= -g_{\mu \nu} f + n\frac{\partial f}{\partial n} \big(U_{\mu} U_{\nu} +g_{\mu \nu} \big) + \frac{\partial f}{\partial \ourG} \big(2 G_{\mu \nu} + g_{\mu \nu} \ourG \big) + E_{\mu \nu}{}^{\lambda} \partial_{\lambda} \frac{\partial f}{\partial \ourG} \,,
\end{align}
where the object $E_{\mu \nu}{}^{\lambda}$ comes from the variations of the bulk term and is defined in 
\cite{Boehmer:2021aji,Jensko:2023lmn} 
\begin{equation} \label{E}
    E^{\mu \nu \lambda} := \frac{2 \ourG}{\partial( g_{\mu \nu,\lambda})} = 2 g^{\rho \mu} g^{\nu \sigma} \Gamma^{\lambda}_{\rho \sigma} -
  2 g^{\lambda (\mu} g^{\nu) \sigma} \Gamma^{\rho}_{\rho \sigma} + g^{\mu \nu} g^{\lambda \rho} 
  \Gamma^{\sigma}_{\sigma \rho} - g^{\mu \nu} g^{\rho \sigma} \Gamma^{\lambda}_{\rho \sigma} \, .
\end{equation}
Thus, the interaction tensor is coordinate dependent and is required to satisfy certain consistency conditions, see Appendix \ref{append_diff}. When these conditions are satisfied, we can use the total covariant conservation law~(\ref{noncons}) to define the current $Q_{\mu}$ which describes the flow of energy-momentum between the scalar field, matter and the interaction term
\begin{equation}
    \nabla^{\mu} T_{\mu \nu}^{(\phi)} =  \frac{\partial f}{\partial \phi} =: Q_{\nu} \, .
\end{equation}
In this case, it follows that
\begin{equation}
    \nabla^{\mu} \big( T^{(\rm fluid)}_{\mu\nu}  + T_{\mu\nu}^{(\mathrm{int})}\big) = - Q_{\nu} \, .
\end{equation}
This will be confirmed for the cosmological equations in the next section, and is a non-trivial result of using coordinates compatible with our diffeomorphism-breaking models. The deeper reason for the consistency of this theory is related to symmetric teleparallelism,  which can be seen as the covariant analogue of our models. A detailed explanation on this topic can be found in~\cite{BeltranJimenez:2022azb,Jensko:2024bee}, but, for our purposes, it suffices that FLRW cosmology is consistent when studied in Cartesian coordinates.

\subsection{Comment on algebraic boundary couplings}
\label{algeb}

Before studying our model, let us comment on the possibility of using the term $\ourB$ instead of $\ourG$ for the algebraic coupling. In the previous work~\cite{Boehmer:2024rqk}, it was indeed suggested to consider
\begin{equation}
    \mathcal{L}_{\textrm{int}} = - \sqrt{-g} f(n,s,\phi, \ourB) \,.
\end{equation}
Such a coupling will generically lead to fourth-order field equations due to the non-linear function of the boundary term $\ourB$, making a dynamical systems analysis complex. In order for the metric equations to remain second order, the following function can be chosen $f=f(n,s,\phi) \ourB$. Using the usual Hubble function $H=\dot{a}/a$ the cosmological field equations\footnote{The cosmological metric is defined in the next section, while the value of the boundary term for this metric can be found in Eq.~(\ref{appendB}).} then read
\begin{align} 
    \label{cosmo1app}
    &3H^2=\kappa\left(\rho+\frac{1}{2} \dot{\phi}^2+V+6 H\dot{\phi}\pdv{f}{\phi}-18 nH^2\pdv{f}{n}\right),\\
    \label{cosmo2app}
    &3H^2+2\dot{H}=-\kappa\Bigl(p+\frac{1}{2} \dot{\phi}^2-V-18 n^2H^2 \pdv[2]{f}{n}+12 H n \dot{\phi } \pdv{f}{n}{\phi}+12 \dot{H} n \pdv{f}{n}-
    2\frac{\dd}{\dd t}\Bigl(\pdv{f}{\phi}\dot{\phi}\Bigr)\Bigr),\\
    &\ddot{\phi}+3H\dot{\phi}+\dv{V}{\phi}+3\left(2\dot{H}+6H^2\right)\pdv{f}{\phi}=0\,, \label{cosmo3app}
\end{align}
where the dot stands for differentiation with respect to cosmological time $t$. However, it is important to remark that these algebraic models are closely related to the derivative coupling models $\mathcal{L}= -\sqrt{-g} f(n,s,\phi) B^{\mu} \partial_{\mu} \phi$ considered in~\cite{Boehmer:2024rqk}. This can be shown by integrating by parts
\begin{align} 
    \label{model_relation}
     -\sqrt{-g} f B^{\mu} \partial_{\mu} \phi &= -\partial_{\mu}\left(\sqrt{-g} f B^{\mu} \phi \right) +  \phi \partial_{\mu}(\sqrt{-g}  B^{\mu}  f) \nonumber \\
    &= \sqrt{-g} \ourB f \phi + \sqrt{-g} \phi B^{\mu} \partial_{\mu} f \, ,
\end{align}
where on the final line we have discarded a total derivative and used $\sqrt{-g} \ourB = \partial_{\mu} (\sqrt{-g}B^{\mu})$.
Making the field redefinition $f(n,s,\phi) \rightarrow -\frac{1}{2} \phi f(n,s,\phi)$ brings the derivative and algebraic interacting Lagrangians into the same form, which then differ only by $-\frac{1}{2} \sqrt{-g} \phi B^{\mu} \partial_{\mu} f$.
In cosmology, the boundary vector has the non-vanishing component $B^{0} = 6 H$, and so this term is
\begin{equation} 
      -\frac{1}{2} \sqrt{-g} \phi B^{\mu} \partial_{\mu} f = -3 a^3 H \phi \left( \pdv{f}{\phi} \dot\phi - 3 H n \pdv{f}{n} \right). \label{eq:condition} 
\end{equation}
It is therefore not too difficult to see that taking the variations leads precisely to the terms that differ between the cosmological equations in the derivative coupling models and our previous equations, up to the aforementioned redefinition of $f$. For example, the constant interaction models studied in the derivative-coupling case~\cite{Boehmer:2024rqk} is equivalent to choosing a function linear in $\phi$ and logarithmic in $n$ for the algebraic case. For this reason, we have chosen to focus on models with non-minimal coupling to the bulk term $\ourG$, ensuring our results are distinct and novel.

\section{Cosmological dynamics}
\label{sec:cosmology}

\subsection{Cosmological field equations}
Let us begin with the homogeneous, isotropic, and spatially flat Friedmann-Lema\^itre-Robertson-Walker (FLRW) metric
\begin{equation}
    \dd s^2=- \dd t^2+a^2(t)\left(\dd x^2+\dd y^2+\dd z^2\right),
\end{equation}
where $a(t)$ is the scale factor.
The bulk term takes the form
\begin{equation} \label{Gbulk}
    \ourG = - 6 H^2 \, ,
\end{equation}
and the cosmological field equations can be written as
\begin{align} \label{cos1}
    3H^2 &= \kappa \big(\rho + \frac{1}{2} \dot{\phi}^2 + V(\phi) + f + 12H^2 f_{,\ourG} \big) \, , \\
    3H^2+ 2 \dot{H} &= -\kappa \Big(p + \frac{1}{2}\dot{\phi}^2 - V(\phi)  - f - 12H^2 f_{,\ourG} - 4\dot{H} f_{,\ourG}  - 4 H \dot{f}_{,\ourG}
    \Big) \, , \label{cos2}
\end{align}
where $\dot{f}_{,\ourG}  = \dd(f_{,\ourG})/\dd t$ implicitly includes derivatives of $n$, $s$, $\phi$ and $H$. Note the equivalence in form with the popular $f(T)$ or $f(Q)$ teleparallel gravity models at the background level\footnote{For instance, our field equations take the same form as non-minimally coupled $f(T,\phi)$ models
\cite{Xu:2012jf,Hohmann:2018rwf,Gonzalez-Espinoza:2020jss,Duchaniya:2022fmc}. However, our function $f$ contains additional couplings to the matter variables $n$ and $s$, which has not been studied before.}, with the torsion scalar $T$ or nonmetricity scalar $Q$ equal to our bulk term~(\ref{Gbulk}). When the non-minimal couplings contained in $f$ are switched off, the full cosmological equations~(\ref{cos1})--(\ref{cos2}) are identical to those studied in both $f(T)$ and $f(Q)$ gravity, e.g.,~\cite{Ghosh:2023amt,Ghosh:2024ueq,Bahamonde:2019shr,Kadam:2022lgq}. For further discussion on the equivalence between these types of geometric theories and our diffeomorphism breaking models, we refer to~\cite{Boehmer:2021aji,BeltranJimenez:2022azb,Jensko:2024bee,Jensko:2023lmn}.

The modified Klein-Gordon equation is simply
\begin{equation}
    \ddot{\phi} + 3H \dot{\phi} + V'(\phi) + f_{,\phi} =0  \, .
\end{equation}
Lastly, the conservation of particle number flux and entropy per particle~(\ref{nscons}) take their usual form~\cite{Boehmer:2015kta}
\begin{equation} \label{nscons2}
    \dot{n} + 3 n H = 0 \, , \quad \dot{s} =0 \, ,
\end{equation}
from which we obtain $n \propto a^{-3}$ and $s=s_0 = \textrm{constant}$. We therefore ignore entropy, which contributes only a constant term and can be absorbed by our function $f$.
From the cosmological equations~(\ref{cos1})--(\ref{nscons2}), a non-trivial calculation reveals that the continuity equation holds and the Brown matter fluid is conserved on FLRW backgrounds
\begin{equation}
    \dot{\rho} + 3H(\rho + p) = 0 \, .
\end{equation}
This also follows from the conservation of the longitudinal part of the matter energy-momentum tensor $U^{\nu} \nabla^{\mu}T_{\mu \nu} =0$ applied to cosmological backgrounds~\cite{Boehmer:2015kta}. Consequently, there are no additional constraints that the model must satisfy, allowing us to choose any function $f$.

The density parameters for matter, the scalar field, and the interaction are defined, respectively, as
\begin{equation} \label{densities}
    \Omega_m = \frac{\kappa \rho}{3H^2}\,, \quad 
    \Omega_{\phi} = \frac{\kappa}{3H^2}\Bigl(\frac{1}{2} \dot{\phi}^2 + V(\phi)\Bigr)\,, \quad 
    \Omega_{\textrm{int}} = \frac{\kappa}{3H^2}\Bigl(f + 12H^2 f_{,\ourG}\Bigr) \,,
\end{equation}
such that $\Omega_m + \Omega_{\phi} + \Omega_{\textrm{int}} = 1$ from~(\ref{cos1}). We will later also refer to the \textit{dark energy density parameter} which is simply $\Omega_{\textrm{DE}}=\Omega_{\phi} + \Omega_{\textrm{int}}$.
For the standard matter sector, we assume a linear equation of state $p = w \rho$, where $w$ takes values between $[-1,1]$. As an additional piece of terminology, it is useful to define the equation of state (EoS) of the scalar field to be
\begin{equation}
     w_{\phi} = \frac{p_\phi}{\rho_{\phi}} = \frac{\frac{1}{2}{\dot{\phi}^2}+ V(\phi)}{\frac{1}{2}{\dot{\phi}^2} - V(\phi)} \,.
\end{equation}
Likewise, we define the equation of state of the interaction term to be
\begin{equation}
    w_{\textrm{int}} = \frac{p_{\textrm{int}}}{\rho_{\textrm{int}}} = 
    -1-4\frac{\dv{t}(H f_{,\ourG} )}{f + 12H^2 f_{,\ourG}}\,.
\end{equation}
Finally, we define the effective dark energy component
\begin{equation} \label{w_DE}
    w_{\textrm{DE}} = \frac{p_{\phi}+p_{\textrm{int}}}{\rho_{\phi} +\rho_{\textrm{int}}} \,.
\end{equation}
Both the scalar field and interaction terms can be interpreted as contributing to dark energy in the early or late Universe. However, we will see that while $w_{\phi}$ is bound between $[-1,1]$, as is standard for quintessence scenarios, the interaction EoS can take on phantom values $w_{\textrm{int}} <-1$. In fact, even the effective dark energy equation of state $w_{\textrm{DE}}$ in~(\ref{w_DE}) can take phantom values too, leading to an array of interesting possibilities. These will be shown in Section~\ref{sec:scalarfield}, where solutions with a transient $w_{\textrm{DE}}<-1$ can give a realistic cosmological scenario.

In order to study concrete solutions, let us partially fix our model by choosing $f(n,s,\phi,\mathbf{G})=\widetilde{f}(n,\phi) \mathbf{G}$. 
The cosmological field equations~(\ref{cos1})--(\ref{cos2}) reduce to
\begin{align} 
    \label{eq:cosmo1}
    &3H^2=\kappa\left(\rho+\frac{1}{2} \dot{\phi}^2+V+6\widetilde{f}H^2\right),\\
    \label{eq:cosmo2}
    &3H^2+2\dot{H}=-\kappa\left(p+\frac{\dot{\phi }^2}{2}-V-6\widetilde{f}H^2-4\widetilde{f}\dot{H}-4H\dot{\phi}\pdv{\widetilde{f}}{\phi}+6H^2n\pdv{\widetilde{f}}{n}\right),\\
    &\ddot{\phi}+3H\dot{\phi}+\dv{V}{\phi}-6H^2\pdv{\widetilde{f}}{\phi}=0\,, 
    \label{eq:cosmo3}
\end{align}
where we have used $\dot{n} = -3Hn$ and $\dot{s}=0$ from~(\ref{nscons2}).
The explicit coupling between the bulk term and the number density $n$ in~(\ref{eq:cosmo2}) is new and has not appeared in the literature.
Note that entropy is constant for adiabatic cosmological fluids at the background level, which means considering a coupling function independent of $s$ is not a restriction. For constant entropy we then have $\rho(n,s)=\rho(n,s_0)=\tilde{\rho}(n)=\tilde{\rho}$. Provided that this function is invertible, we could equally write $n=n(\tilde\rho)$. This means the coupling $\widetilde{f}(n,\phi)$ could be seen as $\widehat{f}(\tilde\rho,\phi)$. As one would expect, this model effectively couples the scalar field and the matter to the geometry. This also shares similarities with the non-minimal couplings studied in $f(T)$ gravity, see~\cite{Gonzalez-Espinoza:2020jss}.

\subsection{Dynamical systems formulation}

Dynamical systems techniques have become one of the standard tools to investigate the background evolution of cosmological models~\cite{Bahamonde:2017ize}. The standard dynamical variables are motivated by the constraint equation~(\ref{eq:cosmo1}) after dividing by $3H^2$. All terms which now appear on the right-hand side are dimensionless, like the cosmological density parameters. However, where possible, one also takes into account the signs of these terms. This motives the variables 
\begin{equation}
    x=\frac{\sqrt{\kappa} \dot{\phi}}{\sqrt{6} H} \,,\quad 
    y=\frac{\sqrt{\kappa V}}{\sqrt{3} H} \geq 0 \,,\quad 
     z=\frac{H^2}{H_0^2+H^2}  \,,\quad
    \sigma=\frac{\sqrt{\kappa \rho}}{\sqrt{3} H} \geq 0 \,,
    \label{eq:variables}
\end{equation}
which allow us to rewrite the Friedmann constraint~(\ref{eq:cosmo1}) as
\begin{equation} 
    \label{fried1}
    1 - x^2 - y^2 - \sigma^2 - 2 \kappa \widetilde{f}(n,\phi)  = 0 \,.
\end{equation}
The variable $z\in [0,1)$ is introduced to deal with models in which the Hubble function cannot be eliminated from the constraint equation~\cite{Boehmer:2022wln}. The constant $H_0$ is an arbitrary value for the Hubble parameter required in the definition of the variable. One can choose $H_0$ to be today's Hubble parameter so that the plane $z=1/2$ corresponds to the current cosmological epoch. The specific form of $z$ can be further motivated as follows: for simplicity, assume either $a(t) \propto t^p$ such that $H = p/t$, or $a(t) \propto \exp(\gamma t)$ such that $H = \gamma$. It follows that $H$ has the following behaviour:
\begin{alignat}{2}
    &H \to 0 \quad &\Leftrightarrow \quad z \to 0 \quad &\Leftrightarrow \quad t \to \infty  \,,  \nonumber \\
    &H \to \infty \quad &\Leftrightarrow \quad z \to 1 \quad &\Leftrightarrow \quad t \to 0 \,,  \label{eq:zlim} \\
    &H = \bar{H} \quad &\Leftrightarrow \quad z=\bar{z} \quad &\Leftrightarrow \quad t=\bar{t} \text{ or } \bar{H}=\gamma   \nonumber   \, ,
\end{alignat}
where the over-barred quantities simply indicate a specific constant value. Consequently, we can interpret these solutions as follows: the $z=1$ plane corresponds to the early Universe, while the $z=0$ plane corresponds to the late Universe. This will be confirmed in the phase space analysis of the following section.

A perfect fluid with linear equation of state and vanishing entropy satisfies $\rho=n^{1+w}$. Let us choose $\widetilde{f}$ to be of the form
\begin{equation} 
    \label{eq:model}
    \widetilde{f}(n,\phi)=\frac{1}{2}\left(c_1\frac{n^{1+w}}{3H_0^2}+c_2\frac{V(\phi)}{3H_0^2}\right),
\end{equation}
for some constants $c_1$ and $c_2$. The model then describes linear couplings between the gravitational bulk term $\ourG$, the matter density $\rho$, and the scalar field potential $V(\phi)$. Similar to~\cite{Copeland:1997et}, we assume an exponential potential of the form 
\begin{align}
    V(\phi)=V_0 \exp(-\lambda\phi)\,, \qquad \text{with}\quad\lambda > 0\,,
\end{align}
which satisfies $\dd V/\dd \phi = -\lambda V$. For this choice, the potential and its derivative can both be expressed in terms of the variable $y$. Other potentials like power-laws further increase the dimensionality of the system, adding an additional layer of complexity which is unhelpful for the current model.

The interaction function given in Eq.~\eqref{eq:model} can be re-written completely in terms of the variables defined in Eq.~\eqref{eq:variables},
\begin{equation} 
    \widetilde{f}(n,\phi)=\frac{1}{2\kappa}\frac{z}{1-z}\left(c_1 \sigma^2+c_2 y^2\right)\,,
\end{equation}
which allows one to re-write Eq.~\eqref{fried1} as
\begin{equation}
    1-x^2-\left(1+c_2\frac{z}{1-z}\right)y^2-\left(1+c_1\frac{z}{1-z}\right)\sigma^2=0\,.
    \label{fried1b}
\end{equation}
Equation~\eqref{fried1b} can be solved for $\sigma$, which means that we have only three independent variables, $x$, $y$, and $z$. We note that while arbitrary powers of the density $\rho$ and potential $V(\phi)$ could have been included in~(\ref{eq:model}), the resulting system will often be too complicated to study. This can be seen in~\cite{Boehmer:2024rqk}, where the model includes an arbitrary power $\alpha$ of the potential term $V(\phi)$ but only special cases of $\alpha$ are studied due to simplicity. Here, for similar reasons, we therefore choose to study only the linear couplings.

Taking the derivative of the dynamical variables with respect to $N=\log a$ and using the cosmological field equations~(\ref{eq:cosmo1})--(\ref{eq:cosmo3}) leads to the following system of equations
\begin{align}
    x'=\,&\frac{(z-1) \left(3 x \left((w-1) \left(x^2-1\right)+(w+1) y^2\right)-\sqrt{6} \lambda  y^2\right)+z(c_1\mathcal{A}+c_2\mathcal{B})}{2 \left(1+z \left(c_1 x^2+\left(c_1-c_2\right) y^2-1\right)\right)}\nonumber\\
    &+\frac{c_1c_2y^2 z^2 \left(3 (w+1) x-\sqrt{6} \lambda (x^2-y^2)\right)-c_2^2\sqrt{6} \lambda  y^4 z^2}{2 (z-1) \left(1+z \left(c_1 x^2+\left(c_1-c_2\right) y^2-1\right)\right)}\,,
    \label{eq:gen1}\\[1ex]
    y'=\,&\frac{y (z-1) \left(3 (w-1) x^2+3 (w+1) \left(y^2-1\right)+\sqrt{6} \lambda  x\right)+z(c_1\mathcal{C}+c_2\mathcal{D})}{2 \left(1+z \left(c_1 x^2+\left(c_1-c_2\right) y^2-1\right)\right)}\nonumber\\
    &+\frac{c_1c_2 y^3 z^2 \left(3(w+1)-2 \sqrt{6} \lambda  x\right)}{2 (z-1) \left(1+z \left(c_1 x^2+\left(c_1-c_2\right) y^2-1\right)\right)}\,,
    \label{eq:gen2}\\[1ex]
    z'=\,&-\frac{3z(z-1) \left(c_1 z-z+1\right)  \left(w \left(x^2+y^2-1\right)-x^2+y^2-1\right)}{1+z \left(c_1 x^2+\left(c_1-c_2\right) y^2-1\right)}\nonumber\\
    &+\frac{c_2y^2 z^2 \left(c_1 z-z+1\right) \left(-3(w+1)+2 \sqrt{6} \lambda  x\right)}{1+z \left(c_1 x^2+\left(c_1-c_2\right) y^2-1\right)}\,,
    \label{eq:gen3}
\end{align}
where
\begin{align}
    \mathcal{A}(x,y)&=- \left(3 (w+1) x^3+3 x \left((w+3) y^2-w-1\right)-\sqrt{6} \lambda y^2 (x^2 + y^2\right)\,,\\[1ex]
    \mathcal{B}(x,y)&=y^2   \left(-3 (w-1) x+2 \sqrt{6} \lambda  x^2-\sqrt{6} \lambda  \left(y^2+1\right)\right)\,,\\[1ex]
    \mathcal{C}(x,y)&=-y  \left(3 (w-1) x^2+3 (w+1) \left(y^2-1\right)+\sqrt{6} \lambda  x(x^2+y^2)\right)\,,\\[1ex]
    \mathcal{D}(x,y)&=-3 y^3  \left(w-\sqrt{6} \lambda  x+1\right) \,.
\end{align}
The matter variable $\sigma$ has been eliminated via the Friedmann constraint~(\ref{fried1b}), rendering the system three-dimensional. 
The system~\eqref{eq:gen1}--\eqref{eq:gen3} generically presents singularities when $z\rightarrow 1$, which is not unexpected given the discussion around Eq.~(\ref{eq:zlim}). However, one may observe that if $c_2=0$, the right-hand sides of Eqs.~\eqref{eq:gen1}--\eqref{eq:gen3} do not present this discontinuity at $z=1$, making the system regular for all $z$. No other choice of $c_1$ or $c_2$ is sufficient to remove these singularities. This is shown explicitly in Appendix~\ref{app:sing}, and will be discussed further in the next section.

Solving Eq.~\eqref{eq:cosmo2} for $\dot{H}$, we can write the deceleration parameter $q=-1-\dot{H}/H^2$ in terms of our dynamical variables as
\begin{multline}
    q=-1-\frac{3\left(\left(c_1-1\right) z+1\right) \left(w \left(x^2+y^2-1\right)-x^2+y^2-1\right)}{2 \left(z
   \left(c_1 x^2+\left(c_1-c_2\right) y^2-1\right)+1\right)}\\-\frac{c_2 y^2 z \left(1+z(c_1-1) \right) \left(-3 w+2 \sqrt{6} \lambda  x-3\right)}{2 (z-1) \left(z
   \left(c_1 x^2+\left(c_1-c_2\right) y^2-1\right)+1\right)}\,.
   \label{eq:q}
\end{multline}
Again, this is singular at $z=1$, which can be seen plainly from the $(z-1)$ factor in the denominator of the final term. Setting $c_2=0$ removes this singularity, with the final term vanishing. One can verify that for $c_1=0$ a factor of $(1-z)$ cancels in the numerator and denominator, again giving a well-defined limit as $z \rightarrow 1$. This behaviour is also shown in Appendix~\ref{app:sing}.
Through $q$ we can find the effective equation of state parameter 
\begin{equation}
    w_{\rm eff} = \frac23 q-\frac13\,.
\end{equation}
It is also useful to write the density parameters~(\ref{densities}) in terms of our dynamical variables
\begin{align}
    \Omega_{\phi} = x^2+y^2 \, , \qquad \Omega_{\textrm{int}} = \frac{z \left(c_1 (1-x^2+y^2) + c_2 y^2 \right)}{1+z(c_1-1)} \, .
\end{align}
One immediately sees that when $z=0$ the interaction density vanishes. This shows that the modification will not directly be relevant at late times $t \rightarrow \infty$ given that $z$ will usually tend to zero there. However, at both early and intermediate times the interaction density will be non-negligible and scale with the parameters $c_1$ and $c_2$. 

\subsection{Fixed points and general properties}

To study the system~\eqref{eq:gen1}--\eqref{eq:gen3}, we now employ dynamical systems theory techniques~\cite{wiggins1990introduction,perko2013differential}. In what follows, we introduce the key concepts which allow us to study the three-dimensional system of autonomous ordinary differential equations. If we consider the coordinates $\vec{x}\in \mathbb{R}^3$ and the map $\vec{f}:\mathbb{R}^3\to \mathbb{R}^3$, the autonomous system can be written in the form
\begin{equation}
    \vec{x}'=\vec{f}(\vec{x}(\tau))\,,\label{eq:dynsys}
\end{equation}
where the prime denotes differentiation with respect to the time variable $\tau\in\mathbb{R}$. In our case, $\tau = N = \log a$. A fixed (critical, or stationary) point of Eq.~\eqref{eq:dynsys} is a point $\vec{x}_0$ such that $\vec{f}(\vec{x}_0)=0$. Linearising the system around $\vec{x}_0$ allows us to classify the fixed points using linear stability theory. The eigenvalues of the Jacobian matrix 
\begin{equation*}
    J =\begin{pmatrix}
    \partial_x f_1 & \partial_y f_1 & \partial_z f_1\\
    \partial_x f_2 & \partial_y f_2 & \partial_z f_2\\
    \partial_x f_3 & \partial_y f_3 & \partial_z f_3
    \end{pmatrix},
\end{equation*}
evaluated at any of the critical points $\vec{x}_0$ determine the linear stability~\cite{wiggins1990introduction} near $\vec{x}_0$. Fixed points are called nonhyperbolic if any of the eigenvalues have zero real part, and different methods must be used to study stability. However, if any of the real non-zero eigenvalues are positive, one can instantly determine that the point is unstable. The only nonhyperbolic point we will encounter will satsify this property, and so methods beyond linear stability theory will not be required.

The form of Eqs.~\eqref{eq:gen1}--\eqref{eq:gen3} is highly complex, including polynomial equations up to quintic order. However, on the $z=0$ plane one finds that the fixed points of the system reduces to the standard quintessence scenario~\cite{Copeland:1997et}, but presented in an alternative set of variables (with different stability properties). New dynamics and new fixed points will therefore only present themselves for $0 < z \leq 1$, largely simplifying the analysis. It follows fairly quickly that the new points with $0<z<1$ must have and $x$-coordinate of $x=0$, leaving two remaining equations which are not difficult to deal with. The special case where $z \rightarrow 1$ will be studied separately for each model, as this limit requires extra care.

The critical points for the system in Eqs.~\eqref{eq:gen1}--\eqref{eq:gen3}, assuming $0 \leq z<1$, are given in Tab.~\ref{tab:generalfixedpoints}. In addition to these, we also find fixed points in the limit $z \to 1$ which are given in Tab.~\ref{tab:projectedfixedpoints}; these are subject to the condition $c_2=0$, for which the dynamical equations~\eqref{eq:gen1}--\eqref{eq:gen3} remain finite at $z = 1$. The corresponding stability analysis and the values of the physical parameters $q$ and $w_{\textrm{eff}}$ for all fixed points are presented in Tab.~\Ref{tab:stabilitygeneral}  Tab.~\ref{tab:stabilityprime} for $0 \leq z <1$ and $z\rightarrow1$ respectively. 

\begin{table}[!htb]
\centering
\begin{tabular}[!thb]{|l|c|c|c|c|c|}
\hline
Point & Coordinates $(x,y,z)$ & $\Omega_m$ & $\Omega_{\phi}$ & $\Omega_{\textrm{int}}$ & Existence\\
\hline \hline
$O$ & $0,\ 0,\ 0$ & $1$ & $0$ & $0$ &  always \\
$A_{-}$& $-1,\ 0,\ 0$ & $0$ & $1$ & $0$ & always \\
$A_{+}$& $1,\ 0,\ 0$ & $0$ & $1$ & $0$ & always \\[1ex]
$B$ & $\displaystyle\sqrt{\frac{3}{2}}\frac{(w+1)}{\lambda},\ \sqrt{\frac{3}{2}}\frac{\sqrt{1-w^2}}{\lambda},\ 0$ 
& $ \displaystyle  1 - \frac{3(1+w)}{\lambda^2}$ & $ \displaystyle \frac{3(1+w)}{\lambda^2}$ & $0$ &$\lambda>\sqrt{3(w+1)}$ \\[2ex]
$C$ & $\displaystyle\frac{\lambda}{\sqrt{6}},\ \sqrt{1-\frac{\lambda^2}{6}},\ 0$
& $0$ & $1$ & $0$
& $0 < \lambda < \sqrt{6}$\\[1ex]
$D$& $\displaystyle0,\ 0,\ \frac{1}{1-c_1}$
& $\infty$ & $0$ & $- \infty$ & $c_1<0$\\[1ex]
$E$& $\displaystyle0,\ \frac{1}{\sqrt{2}},\ \frac{1}{1+c_2}$
& $0$ & $\displaystyle \frac{1}{2}$ & $\displaystyle  \frac{1}{2}$ & $c_2>0$\\[2ex]
\hline
\end{tabular}
\caption{Critical points for Eqs.~\eqref{eq:gen1}--\eqref{eq:gen3} and their existence conditions, assuming $-1 < w \leq 1$ and $\lambda>0$.}
\label{tab:generalfixedpoints}
\end{table}

\begin{table}[!thb]
\centering
\begin{tabular}[t]{|l|c|c|c|c|c|}
\hline
Point & Coordinates $(x,y,z)$ & $\Omega_m$ & $\Omega_{\phi}$ & $\Omega_{\textrm{int}}$ & Existence\\
\hline
$A'_{-}$ & $-1,\ 0,\ 1$ & $0$ & $1$ & $0$  & always  \\
$A'_{+}$ & $1,\ 0,\ 1$ &  $0$ & $1$ & $0$ & always\\[1ex]
$C'$ & $\displaystyle\frac{\lambda}{\sqrt{6}},\ \sqrt{1-\frac{\lambda^2}{6}},\ 1$  & $0$ & $1$ & $0$ & 
 $0 < \lambda < \sqrt{6}$\\[2ex]
\hline
\end{tabular}
\caption{Fixed points for system~\eqref{eq:gen1}--\eqref{eq:gen3} in the limit $z\to 1$, subject to the condition that $c_2=0$.}
\label{tab:projectedfixedpoints}
\end{table}

Let us begin by noting that the fixed points $O$, $A_{\pm}$, $B$, and $C$, which lie on the plane $z=0$, are analogous to the standard exponential potential fixed points studied by Copeland et al.~\cite{Copeland:1997et}. For all of these points, the interaction does not play a role and its density parameter vanishes $\Omega_{\textrm{int}}=0$. However, for our model $A_{-}$ and $A_{+}$ are always saddle points as opposed to behaving as unstable nodes for some values of $\lambda$; this follows from the fact that we use the variable $z$, which now differentiates between the early and late Universe. The other three points $O$, $B$, and $C$ exhibit the same behaviour described in~\cite{Copeland:1997et}. The origin $O$ corresponds to the matter-dominated saddle with $\Omega_m=1$ and $w_{\textrm{eff}} = w$. At points $A_{\pm}$, the universe is dominated by the kinetic energy of the scalar field $x^2 = 1$ and the effective equation of state parameter takes the value of stiff matter ($w=1$). Similarly to~\cite{Copeland:1997et}, the fixed point $B$ is a matter \emph{scaling solution} since the effective equation of state parameter $w_{\rm eff}$ matches that of the matter component, $w$. Point $C$ lies on the unit half-circle at $z=0$ and is scalar-field-dominated $\Omega_{\phi}=1$. For $\lambda<\sqrt{2}$, the deceleration parameter is negative $q<0$ and this point represents an accelerating solution.

On the $z=1$ plane, subject to the requirement that $c_2 = 0$, we have three critical points $A'_{-}$, $A'_{+}$ and $C'$. These share the same properties as their unprimed counterparts on the $z=0$ plane, with the same values of $q$ and $w_{\textrm{eff}}$ and the same stability classification. It is therefore straightforwrd to interpret these as `early-time' analogues of $A_{-}$, $A_{+}$ and $C$ respectively. 

Additionally, we find two new points which depend on the coupling constants $c_1$ and $c_2$.
Point $D$ exists only when $c_1<0$, and is nonhyperbolic as one of its eigenvalues is zero (contains a centre manifold). However, since the other two eigenvalues are positive and negative, the point also contains a stable and unstable manifold~\cite{perko2013differential}; it follows that point $D$ always displays unstable saddle-like behaviour, irrespective of the stability of its centre manifold. The effective equation of state at this point is $w_{\rm eff}=-1$, representing a de Sitter solution. We will later see that it can be interpreted as an early-time inflationary point, which follows from the fact that the matter and dark energy density parameters diverge to plus and minus infinity here, $\Omega_{m} \rightarrow + \infty$ and $\Omega_{\textrm{DE}} \rightarrow - \infty$. The physical implications of trajectories approaching/originating from point $D$ will be discussed in detail in the next section.
On the other hand, the fixed point $E$ exists only when $c_2>0$ and is always a saddle. This point lies on the boundary of the physical phase space and is again a de Sitter solution, as can be seen from $q=w_{\rm eff}=-1$. Notably, the point is dark energy dominated $\Omega_{\textrm{DE}}=1$, with an equal split between the canonical scalar field part $\Omega_{\phi}=1/2$ and the modified gravity interaction part $\Omega_{\textrm{int}}=1/2$. This point will be relevant for the cosmological dynamics at intermediate times, playing a key role in the transient phantom behaviour of the models with $c_2 > 0$.

\begin{table}[!htb]
\centering
\begin{tabular}[t]{|l|c|l|c|c|}
\hline
Point & Eigenvalues & Stability & $q$ & $w_{\rm eff}$\\
\hline\hline
$O$ & $\displaystyle\frac{3}{2} (w\pm1),\ -3 (w+1)$ & saddle & $\frac{1}{2}(1+3w)$ & $w$ \\[1ex]
$A_{-}$ & $\displaystyle -6,\ 3(1-w),\ 3+\sqrt{\frac{3}{2}} \lambda$ & saddle & 2 & 1 \\[2ex]
$A_{+}$& $\displaystyle -6,\ 3(1-w),\ 3-\sqrt{\frac{3}{2}} \lambda$ & saddle & 2 & 1\\[2ex]
$B$ & $\displaystyle -3 (w+1),\ \frac{3}{4 \lambda } \left[\lambda  (w-1)\pm\Delta\right]$ &
\multirow{6}{14em}{%
stable if $-1<w\leq-7/9$ \\ and $\lambda>\sqrt{3(w+1)}$\\[2ex]
stable if $-7/9<w<1$ and\\ $\sqrt{3(w+1)}< \lambda\leq2\sqrt{\frac{6(w+1)^2}{9w+7}}$ \\[2ex]
stable spiral if $-7/9<w<1$ \\and $\lambda>2\sqrt{\frac{6(w+1)^2}{9w+7}}$
} & $\frac{1}{2}(1+3w)$ & $w$ \\[10em]
$C$ & $\displaystyle -\lambda ^2,\ \lambda ^2-3(w+1),\ \frac{1}{2} \left(\lambda ^2-6\right)$ & 
\multirow{2}{14em}{stable if $0<\lambda<\sqrt{3(w+1)}$ \\[1ex]
saddle if $\sqrt{3(w+1)}<\lambda<\sqrt{6}$} & $-1+\frac{\lambda^2}{2}$ & $-1+\frac{\lambda^2}{3}$ \\[4em]
$D$ & $-3,\ 0,\ 3(1+w)$ & (nonhyperbolic) unstable & $-1$ & $-1$\\[1ex]
$E$ & $\displaystyle-3(w+1), \
\frac{1}{2} \left(-3\pm\sqrt{12 \lambda ^2+9}\right)$ & saddle & $-1$ & $-1$\\[1ex]
\hline
\end{tabular}
\caption{Stability of critical points for Eqs.~\eqref{eq:gen1}--\eqref{eq:gen3} and values of the deceleration parameter $q$ and $w_{\rm eff}$ at the fixed points. Here $\Delta=\sqrt{(w-1)\left(\lambda ^2 (9 w+7)-24 (w+1)^2\right)}$.}
\label{tab:stabilitygeneral}
\end{table}

\begin{table}[!htb]
\centering
\begin{tabular}[t]{|l|c|l|c|c|}
\hline
Point & Eigenvalues & Stability & $q$ & $w_{\rm eff}$\\
\hline\hline
$A'_{-}$ & $\displaystyle 6,\ -3(1+w),\ 3+\sqrt{3/2} \lambda$ & saddle & 2  & 1  \\[1ex]
$A'_{+}$ & $\displaystyle 6,\ -3(1+w),\ 3-\sqrt{3/2} \lambda$ 
& saddle & 2 & 1 \\[1ex]
$C'$ & $\displaystyle -\lambda ^2,\ \lambda ^2-3(w+1),\ \frac{1}{2} \left(\lambda ^2-6\right)$ & 
\multirow{2}{14em}{stable if $0<\lambda<\sqrt{3(w+1)}$ \\[1ex]
saddle if $\sqrt{3(w+1)}<\lambda<\sqrt{6}$} & $-1+\frac{\lambda^2}{2}$ & $-1+\frac{\lambda^2}{3}$ \\[4ex]
\hline
\end{tabular}
\caption{Stability of critical points as $z \to 1$ for Eqs.~\eqref{eq:gen1}--\eqref{eq:gen3}, as in Tab.~\ref{tab:projectedfixedpoints}, and values of the deceleration parameter $q$ and $w_{\rm eff}$ at the fixed points. Note that this is subject to the condition $c_2=0$.}
\label{tab:stabilityprime}
\end{table}

\clearpage

\section{Phase space analysis}
\label{sec:phase}

We are now ready to study concrete models with various coupling functions $\widetilde{f}$. The first model modifies the standard quintessence scenario with the inclusion of the early-time de Sitter point $D$, see Tab.~\ref{tab:stabilitygeneral}; this is obtained by setting $c_2 = 0$ along with $c_1 < 0$, which corresponds to coupling the energy density $\rho$ to $\ourG$. The second model instead affects the intermediate-to-late-time dynamics with the inclusion of the saddle point $E$ for $c_1=0$ and $c_2 >0$, see Tab.~\ref{tab:stabilitygeneral}; this model describes a pure scalar field coupling between the scalar field $\phi$ and geometrical pseudo-scalar $\ourG$. We then briefly discuss the class of `hybrid models' with $c_1 < 0$ and $c_2 >0$, in order for both de Sitter points $D$ and $E$ to exist simultaneously within the physical phase space. This last model represents the most complex phenomenology and can contain all critical points listed in Tab.~\ref{tab:generalfixedpoints}. To investigate all of these models, we first closely study their physical phase spaces, as determined by the Friedmann constraint. We then perform numerical analyses to calculate trajectories through the phase space, studying the evolution of the physical parameters along these orbits.

\subsection{Energy density couplings}

For the first set of models we set $c_2=0$. The couplings to the scalar field are switched off and the interaction of the model reduces to
\begin{align}
    \widetilde{f}(n,\phi)\equiv\widetilde{f}(n)=c_1\frac{1}{2}\frac{n^{1+w}}{3H_0^2}=\frac{c_1}{2\kappa }\frac{z}{1-z}\sigma^2\,.
\end{align}
Given that $\rho \propto n^{1+w}$, we are effectively dealing with a coupling of the form $f(n,s,\phi,\ourG)\propto \rho\, \vec{G}$. Interestingly, this is the type of coupling one might have proposed even without input from dynamical systems considerations: one simply couples the matter energy density non-minimally to a quantity related to curvature (in this case, the bulk term). If the pure-gravitational part of the action is described by $\ourG$, then these types of non-minimal couplings would naturally arise in an effective field theory approach.
Moreover, as matter is dominant at earlier times of the cosmological evolution, such couplings should predominantly affect the early-time Universe.

\subsubsection{Physical phase space}

For $c_2=0$, the Friedmann constraint~(\ref{fried1b}) simplifies to
\begin{align} 
    x^2 + y^2 + \left(1 + c_1\frac{z}{1-z}\right)\sigma^2 = 1 \,,
\end{align}
which allows us to write 
\begin{align} 
    \sigma^2 = \frac{1-x^2-y^2}{1 + c_1z/(1-z)}\,.
    \label{constraintc1}
\end{align}
Additionally, we impose the physical conditions
\begin{align} 
    \sigma \geq 0\,, \quad 0 \leq z < 1\,, \quad y \geq 0 \,.
\end{align}
These conditions ensure that energy density is non-negative, that the scalar field's potential is positive, and that the Hubble function is positive and bounded by infinity. 

\begin{figure}[!htb]
	\centering
	\begin{subfigure}{0.36\textwidth}
    \includegraphics[width=\textwidth]{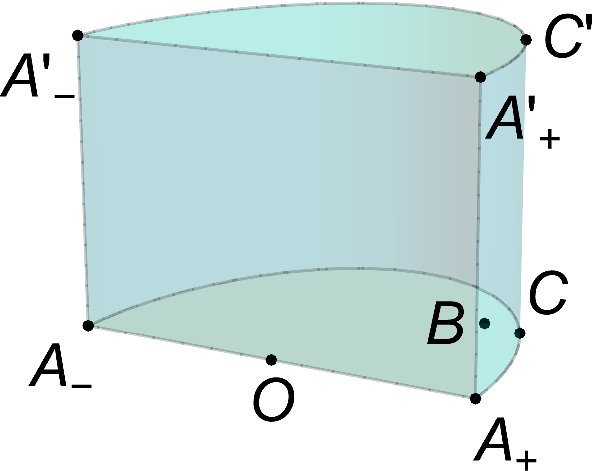}
            \subcaption{$c_1 >0$}
        \label{fig:bound1}
        \end{subfigure}
        \hspace{6mm}
	\begin{subfigure}{0.47\textwidth}
            \includegraphics[width=\textwidth]{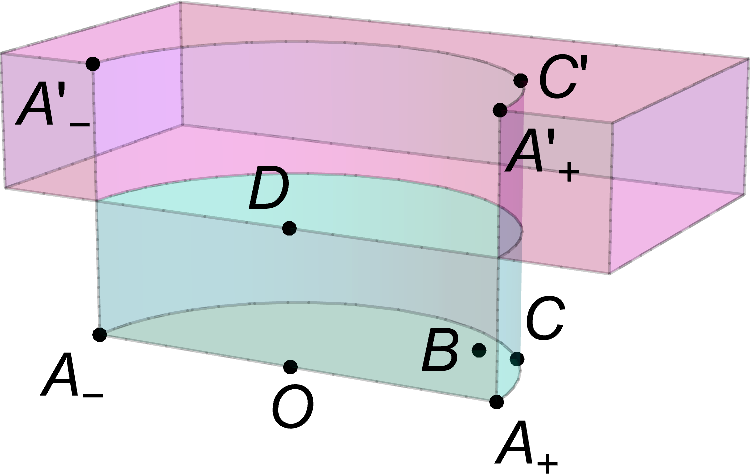}
            \subcaption{$c_1 < 0$}
            \label{fig:bound2}
        \end{subfigure} 
        \caption{Physical phase space and fixed points for the $c_2=0$ model, given by Eqs.~\eqref{eq:gen1}--\eqref{eq:gen3}. The fixed point $D$ only exists for $c_1<0$.} \label{fig:boundM1}
\end{figure}

For $c_1 \geq 0$, the positivity of the matter density parameter $\sigma\geq 0$ implies the simple condition $1-x^2-y^2 \geq 0$. This leads to a phase space defined by a half cylinder of unit height, illustrated in Fig.~\ref{fig:bound1}. Trajectories will therefore start near the $z=1$ plane, corresponding to the early-time Universe $t \to 0$, and terminate on the $z=0$ plane, where $t \to \infty$. 
This case is simply a projection of the Copeland et. al. model~\cite{Copeland:1997et} and does not exhibit any novel fixed points\footnote{We note that in the $z \rightarrow1$ limit, the Friedmann constraint breaks down and one cannot assume the phase space is confined to the half unit circle. It is therefore possible that fixed points at infinity can exist in this limit, but this would only be relevant at early times.}.

For negative values of $c_1$, the denominator of the right-hand side of Eq.~\eqref{constraintc1} can change sign when
\begin{align} 
    z_0 = \frac{1}{1-c_1} \,.
\end{align}
Thus, when $c_1<0$, one has $0 < z < z_0$. Note that the sign change of the denominator on the right-hand side of Eq.~\eqref{constraintc1} can only be compensated by a sign change in the numerator, which means that we must require $1-x^2-y^2<0$. Therefore, the phase space splits into two regions. The upper region, satisfying $1-x^2-y^2<0$, is no longer compact and contains the points $A'_{-}$, $A'_{+}$ and $C'$ on the $z=1$ plane. From Tab.~\ref{tab:stabilitygeneral}, we notice that none of these points are unstable repellers, indicating that trajectories originate from infinity. The lower region, satisfying $1-x^2-y^2>0$, contains the new de Sitter point $D$ at $z=z_0$. The physical phase space is illustrated in Fig.~\ref{fig:bound2}, with the two regions intersecting at $z=z_0$ and $x^2+y^2=1$.

From the Friedmann constraint, one finds that the matter density parameter $\sigma$ is indeterminate on the boundary between the two distinct regions $z=z_0$. Consequently, we consider those regions to be physically distinct. We therefore choose to confine our analysis to the lower region defined by the half cylinder of height $z_0=1/(1-c_1)$. It is interesting to note that the Friedmann constraint remains finite at $z=z_0$, and it may be possible to extract information about trajectories passing between both regions on the semi-circle defined by $x^2+y^2=1$ with $y\geq0$ and $z=z_0$. Provided trajectories approach along the $x^2+y^2=1$ plane, where the density parameter vanishes $\sigma=0$, the system should remain well-behaved. For instance, along the orbit $A'_{-} \rightarrow A_{-}$ the matter density parameter remains zero. However, the upper phase space is non-compact and requires further asymptotic analysis. While this might of some mathematical interest, we do not anticipate this to yield physically relevant results. On the other hand, trajectories in the lower region can originate from point $D$ at $z=z_0$, which may be interpreted as an early-time\footnote{Note that point $D$ is an unstable centre, as shown in Table~\ref{tab:stabilitygeneral}. However, $D$ can act as a source if we only consider the physical region $0\leq z < z_0 $.} inflationary point where $q \rightarrow -1$ and $\Omega_m \rightarrow \infty$. We will study this last scenario in the following numerical analysis.

\subsubsection{Evolution and phase plots}

\begin{figure}[!htb]
    \centering
    \begin{subfigure}{0.8\textwidth}
    \includegraphics[width=\textwidth]{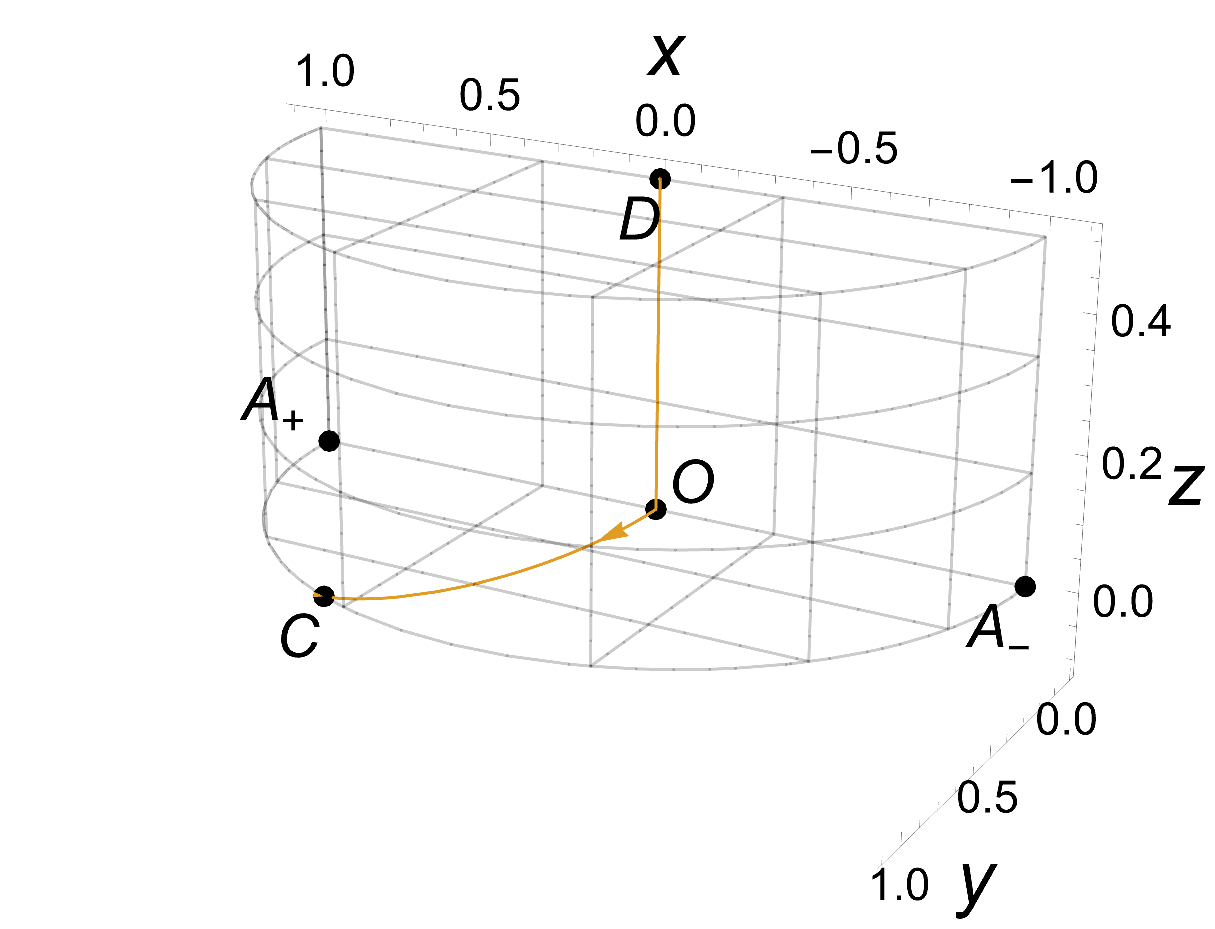}
    \subcaption{Phase space plot highlighting the trajectory $D \to O \to C$.}
    \label{fig:2a}
    \end{subfigure}\\\vspace{.5cm}
    \begin{subfigure}{0.7\textwidth}
    \includegraphics[width=\textwidth]{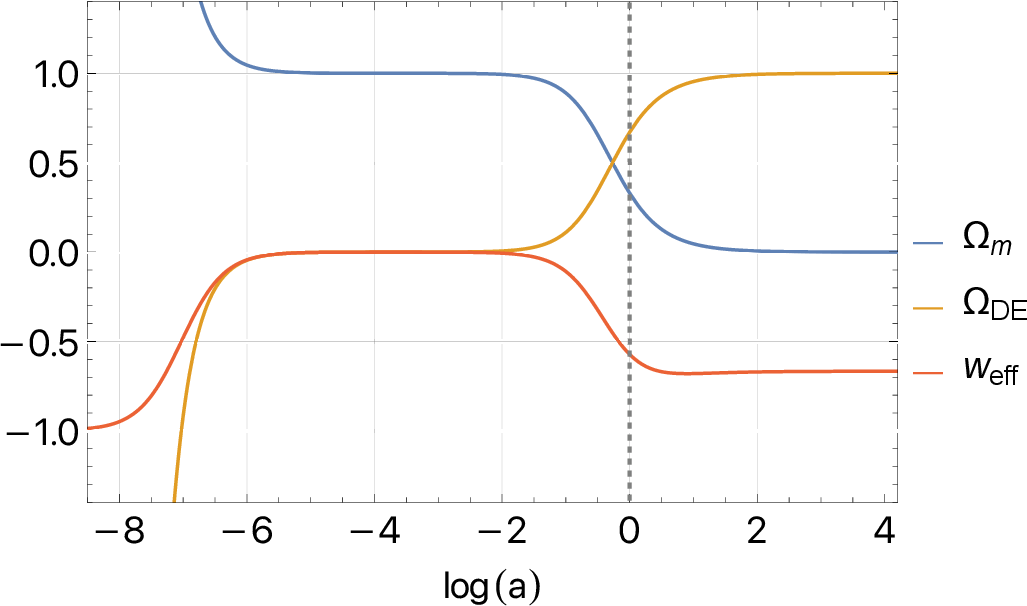}
    \subcaption{Evolution of the physical quantities (matter density $\Omega_m$, dark energy density $\Omega_{\textrm{DE}}$ and effective equation of state $w_{\rm eff}$) for the trajectory shown above.}
    \label{fig:2b}
    \end{subfigure} 
    \caption{Parameter values are $\lambda=1$, $c_1 =-1$, $c_2=0$. As initial conditions we set the value of today's matter density parameter to $\Omega_{m} \approx 0.33$, indicated by the dashed line at $N=0$.}
    \label{fig:2}
\end{figure}

We now continue to assume $c_2=0$ and $c_1 <0$ and restrict our attention to the physical region $0\leq z < z_0$, such that the inflationary point $D$ acts as the early-time repeller of the physical phase space. Trajectories following the heteroclinic orbit $D \rightarrow O \rightarrow C$ have a late-time evolution analogous to the standard minimally coupled quintessence scenario, with the matter couplings $\rho\, \ourG$ only modifying the early-time dynamics. The phase space and orbit for such a trajectory is given in Fig.~\ref{fig:2a}, and the corresponding evolution of the physical parameters is plotted in Fig.~\ref{fig:2b}. 

The most interesting feature displayed by the model is that the effective equation of state starts around $w_{\rm eff} =-1$ close to the de Sitter point $D$, then evolves towards a matter dominate state, $w_{\rm eff} = 0$, and ends at the scalar field potential point $C$ with $w_{\rm eff} = -1+\lambda^2/3 = -2/3$. Such a model allows for early and late-time accelerated expansion without the introduction of additional fields. 
However, it is also possible to reinterpret and express the non-minimal coupling as additional field content. It should be noted that the density parameters $\Omega_m$ and $\Omega_{\rm DE}$ diverge at early times when approaching the point $D$, as is evident from Tab.~\ref{tab:generalfixedpoints}. With the inclusion of a radiation component, such a model would be consistent with standard cosmological observations for certain parameter choices. It would therefore be particularly interesting to study inflationary scenarios in more detail for these energy density coupling models.

\subsection{Scalar field couplings}
\label{sec:scalarfield}

We now turn to the case with $c_1=0$. The fixed point $D$ will no longer be present in the phase space, being replaced by the new saddle point $E$ for $c_2>0$, and the interaction term reduces to
\begin{align}
    \widetilde{f}(n,\phi)\equiv\widetilde{f}(\phi)=c_2\frac{1}{2}\frac{V(\phi)}{3H_0^2}=\frac{c_2}{2\kappa }\frac{z}{1-z}y^2\,.
    \label{coup2}
\end{align}
Hence, the coupling is of the type $f(n,s,\phi,\ourG)\propto \phi\, \vec{G}$. This is analogous to the previously studied cases in teleparallel gravity where the torsion scalar is non-minimally coupled to $\phi$~\cite{Bamba:2013jqa,Bahamonde:2015hza}. The background dynamics of this model is therefore a subset of the $f(T,\phi)$ theories~\cite{Otalora:2013tba,Otalora:2013dsa}. However, our different choice of variables will lead to a unique presentation with some particularly interesting properties. 

\subsubsection{Physical phase space}

\begin{figure}[!htb]
    \centering
    \begin{subfigure}{0.25\textwidth}
    \includegraphics[width=\textwidth]{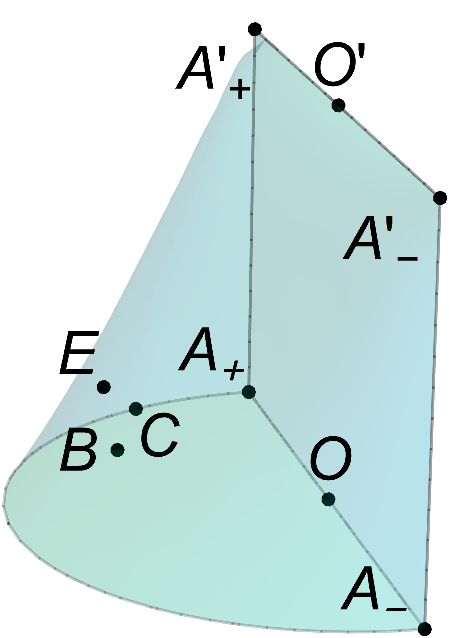}
    \subcaption{$c_2  >0$}
    \label{fig:bound3}
    \end{subfigure}
    \hspace{2cm}
    \begin{subfigure}{0.35\textwidth}
    \includegraphics[width=\textwidth]{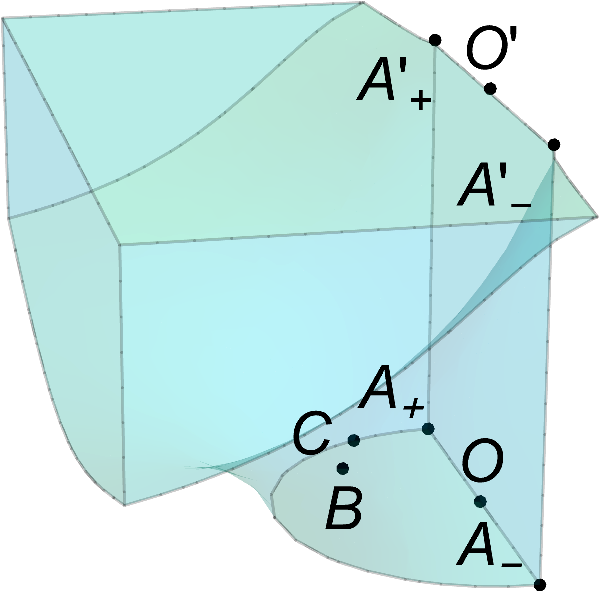}
    \subcaption{$c_2 < 0$}
    \label{fig:bound4}
    \end{subfigure} 
    \caption{Physical phase space and fixed points for the $c_1=0$ model, given by Eqs.~\eqref{eq:gen1}--\eqref{eq:gen3}. The fixed point $E$ only exists for $c_2>0$.}
    \label{fig:boundM2}
\end{figure}

Let us now set $c_1=0$ in Eq.~\eqref{fried1b}, which gives
\begin{align} 
    \sigma^2 = 1 - x^2 - \left(1+c_2\frac{z}{1-z}\right)y^{2}\,.
\end{align}
Along with the physical conditions $\sigma \geq 0$, $ 0 \leq z < 1$ and $y \geq 0$, we obtain the constraint
\begin{align} 
    1 - x^2 - \left(1 + c_2 \frac{z}{1-z}\right)y^2  \geq 0 \,.
    \label{constraintc2}
\end{align}
Therefore, the constant height $z=z_0$ cross-sections of the physical phase space are ellipses if $c_2>0$, see Fig.~\ref{fig:bound3}. When $c_2<0$, the phase space is no longer compact as $c_2 y^2 z/(1-z) \leq 0$, which means $x$, $y$ become unbounded, see Fig.~\ref{fig:bound4}. Physically, this corresponds to the modified gravity component $\Omega_{\textrm{DE}}$ taking negative values, which follows from the Friedmann constraint $\Omega_{m} + \Omega_{\textrm{DE}} = 1$ with $\Omega_{m} >1$. We will not discuss this case further due to the additional complications in dealing with an unbounded phase space and having an unbounded matter variable. Returning to $c_2>0$, we define the eccentricity 
\begin{align}
    e=\sqrt{\frac{c_2 z_0}{1+(c_2-1)z_0}} \,.
\end{align}
These ellipses become narrower ($e$ increases) when either $z_0$ or $c_2$ increase. For $z_0=0$ or $c_2=0$ the ellipse degenerates to the unit circle with $e=0$. For $c_2 \to \infty$ (or $z \to 1$ in~(\ref{constraintc2})), it degenerates to a line, $e \to \infty$. Physically, we expect all trajectories to start on the line $x=0$ and $z \to 1$, which corresponds to the early Universe $t \to 0$, and terminate on the $z=0$ plane, which corresponds to $t \to \infty$, see Fig.~\ref{fig:bound3}. The physical phase spaces for both signs of the parameter $c_2$ are given in Fig.~\ref{fig:boundM2}, where the fixed points of each of model are also shown for completeness.

We now focus on the case where $c_1=0$ and $c_2 > 0$, where the phase space is compact and the fixed point $E$ plays a role at intermediate times. As was shown previously, the $z \to 1 $ line is singular. However, under suitable conditions, one can obtain limits along this line if $y$ approaches zero sufficiently fast. In Appendix~\ref{app:sing}, we discuss these limits in more detail and justify trajectories with $y \to 0$ as $z \to 1$. Consequently, we find a new set of `singular' points on the $z = 1$ line, which behave like critical points when approached from the $y=0$ plane, given in Tab.~\ref{tab:limits}. Note that these points differ slightly from those given in the previous model in Tab.~\ref{tab:stabilityprime}. The eigenvalues of these new points are easily computed, provided that one evaluates the Jacobian by setting $y \to 0$ before taking the limit $z \to 1$. Again, this assumption is justified for physical trajectories in Appendix~\ref{app:sing}.

\begin{table}[hbt!]
\centering
\begin{tabular}[t]{|l|c|c|c|c|c|}
\hline
Point & $(x,y,z)$ & Eigenvalues & Stability & $q$ & Existence\\
\hline
$A'_{-}$ & $-1,\ 0,\ 1$ & $3(1-w), \, 6, \, 3+\sqrt{\frac{3}{2}} \lambda$  & unstable  & 2 & always  \\
$A'_{+}$ & $1,\ 0,\ 1$ & $3(1-w), \, 6, \, 3 - \sqrt{\frac{3}{2}} \lambda$ & 
\multirow{2}{9em}{saddle if $\lambda>\sqrt{6}$\\ unstable if $\lambda<\sqrt{6}$}
 & 2 & always\\[3ex]
$O'$  & $0,\ 0,\ 1$ & $-\frac{3}{2}(1-w),  \, 3(1+w),  \, \frac{3}{2}(1+w)$ & saddle if $-1 < w < 1$& $\frac{1}{2}(1+3w)$
& always\\[2ex]
\hline
\end{tabular}
\caption{Singular points and stability for the $c_1=0$ model in the limit $z\to 1$ and $-1<w\leq 1$, $\lambda >0$.}
\label{tab:limits}
\end{table}

Contrary to the case given in Tab.~\ref{tab:stabilityprime},
where it was assumed that $c_2=0$ along the $z=1$ line, the point $A'_{-}$ now acts like an early-time repeller. We also find the additional point $O'$, which is a matter-dominated saddle. Moreover, the repeller behaviour of the singular point $A'_{-}$ is in fact necessary from a mathematical viewpoint, because unlike the $c_2=0$ case studied in the previous section, the phase space for this model is compact. Consequently, trajectories must originate from one of the finite fixed points, and the stability analysis presented in Tab.~\ref{tab:limits} is therefore consistent with these facts. 

\subsubsection{Evolution plots}

\begin{figure}[hbt!]
	\centering
	\begin{subfigure}{0.8\textwidth}
	    \includegraphics[width=\textwidth]{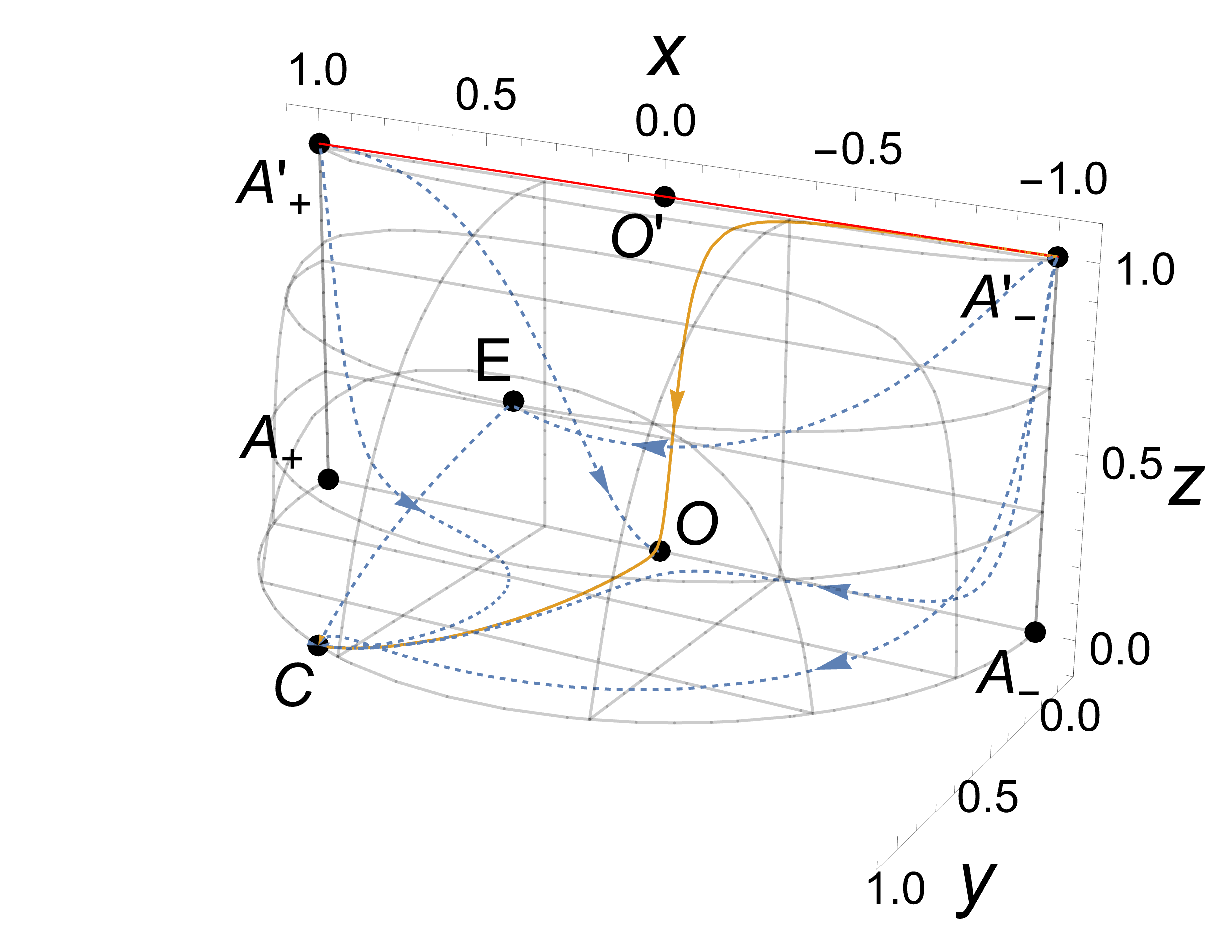}
            \subcaption{
            The dashed blue trajectories are possible trajectories. The orange trajectory represents a physically relevant trajectory starting in a neighbourhood of $A'_{-}$. The solid red line denotes the limit $z\to 1$ which is discontinuous.}
        \label{fig:4a}
        \end{subfigure}\\\vspace{.5cm}
	\begin{subfigure}{0.8\textwidth}
            \includegraphics[width=\textwidth]{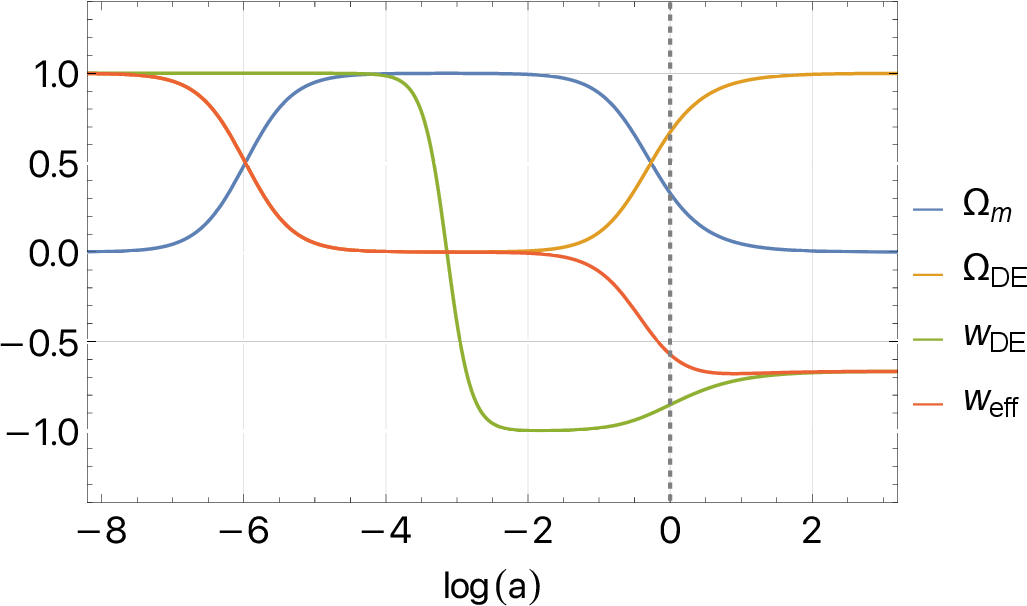}
            \subcaption{Evolution of the physical quantities $\Omega_m$, $\Omega_{\textrm{DE}}$, $w_{\rm DE}$ and $w_{\rm eff}$ for the yellow trajectory shown above.}
            \label{fig:4b}
        \end{subfigure}
        \caption{Here $w=0$, $c_1=0,c_2=1$ and $\lambda=1$, $k=1/2$. Today's value has been taken at $\Omega_m \approx 0.33$} \label{fig:4}
\end{figure}

In Fig.~\ref{fig:4a}, the phase plot for the $c_1=0$ and $c_2 =1 $ model is shown, highlighting a range of possible trajectories within the physical phase space originating from either $A'_{-}$ or $A'_{+}$. The orange trajectory follows the heteroclinic orbit $A'_{-} \rightarrow O' \rightarrow O \rightarrow C$ and gives rise to a standard quintessence scenario evolution, see Fig. \ref{fig:4b}. For $c_2=1$, the new saddle point $E$ has coordinates $(0,1/\sqrt{2},1/2)$ and therefore can potentially be relevant in the current-universe, where $z=1/2$. However, for these types of values, $E$ does not lie close to the heteroclinic orbit (required for periods of matter domination); consequently, for realistic initial conditions and this value of $c_2$, the evolution is largely unmodified by the presence of the new non-minimal couplings and the fixed point $E$.

\clearpage

\begin{figure}[!htb]
    \centering
    \begin{subfigure}{0.8\textwidth}
    \includegraphics[width=\textwidth]{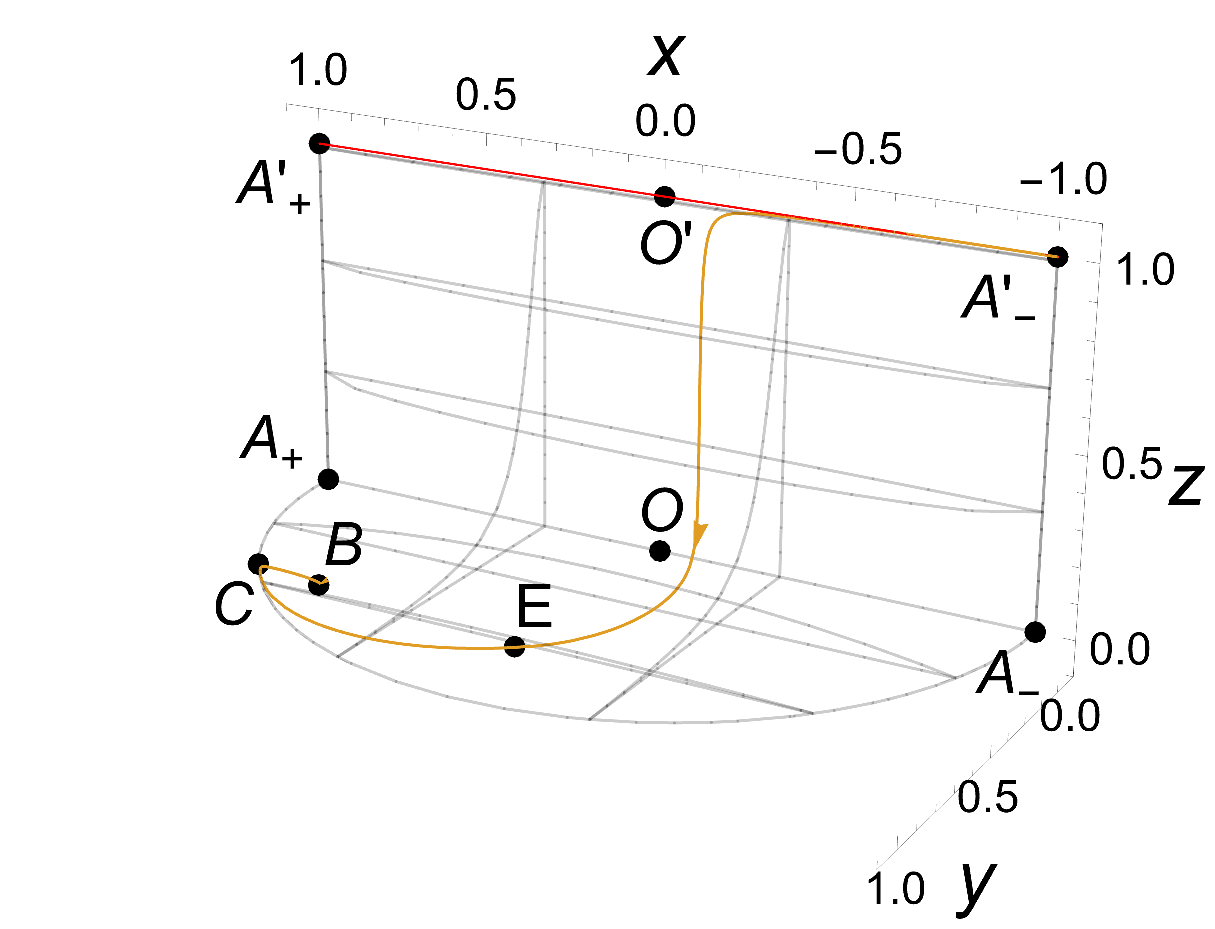}
    \subcaption{The orange trajectory represents a physically relevant trajectory starting in a neighbourhood of $A'_{-}$. The solid red line denotes the limit $z\to 1$ which is discontinuous.}
    \label{fig:im2a}
    \end{subfigure}\\\vspace{.5cm}
    \begin{subfigure}{0.7\textwidth}
    \includegraphics[width=\textwidth]{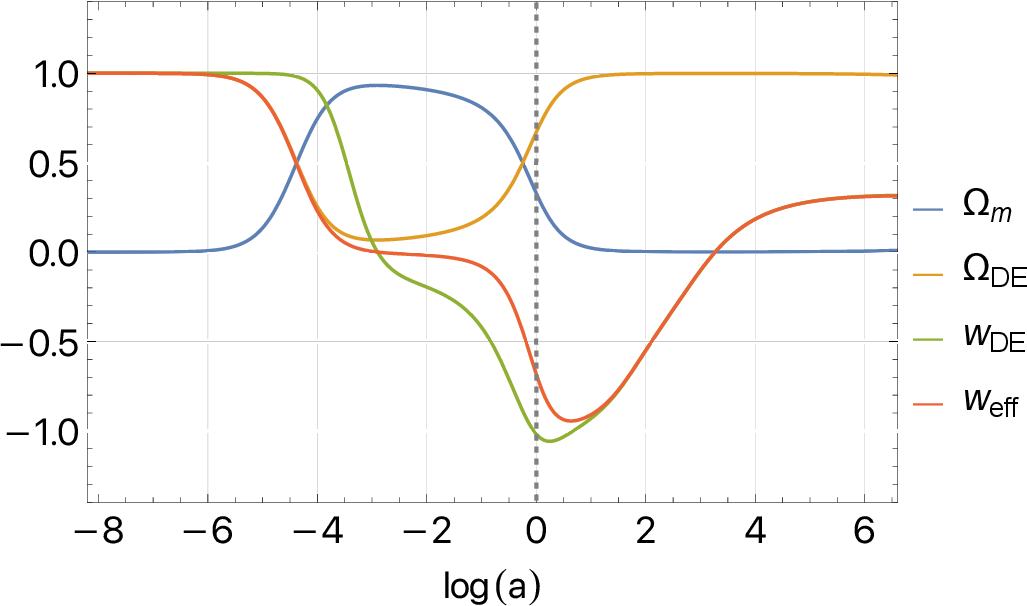}
    \subcaption{Evolution of the physical quantities $\Omega_m$, $\Omega_{\textrm{DE}}$, $w_{\rm DE}$ and $w_{\rm eff}$ for the yellow trajectory shown above.}
    \label{fig:im2b}
    \end{subfigure} 
    \caption{Here $w=0$, $c_1=0$, $c_2=94$ and $\lambda=2$.}
    \label{fig:im2}
\end{figure}

Alternatively, a larger choice of $c_2$ causes the new critical point $E$ to take smaller $z$-values and impact the late-time behaviour. This is demonstrated in Fig.~\ref{fig:im2a} where the value $c_2 = 94$ has been chosen, along with a different value of $\lambda$. In this case, the physical phase space is flattened and the trajectory approaches the new saddle point $E$ before continuing to the standard saddle $C$ and the late-time matter-tracking stable point $B$. The evolution of physical parameters, displayed in Fig.~\ref{fig:im2b}, shows a period of matter domination followed by accelerated expansion at current times. What is particularly interesting is that the dark energy equation of state crosses the phantom divide $w_{\textrm{DE}} <-1$, and at the present $\log(a)=0$ is slightly less than minus one; this is a direct consequence of the presence of the new saddle point $E$. At late times, the trajectory slowly approaches the matter-tracking point $B$ and $w_{\textrm{eff}} \rightarrow 0$, though this is not visible on the plot in Fig.~\ref{fig:im2b}.

For different initial conditions, both the dark energy and effective equation of state can take phantom values too. This phantom crossing behaviour has also been observed in other similar non-minimally coupled models, see for instance~\cite{Otalora:2013tba}. The range of complex dynamical behaviours exhibited just in this simple scalar field coupled model shows the large scope of potential phenomenology allowed in our formulation. It would therefore be especially interesting to use observational data to constrain the free parameters of the model.

\subsection{Hybrid models with matter and scalar field couplings}

\begin{figure}[!htb]
    \begin{subfigure}{0.49\textwidth}
        \includegraphics[width=.9\textwidth]{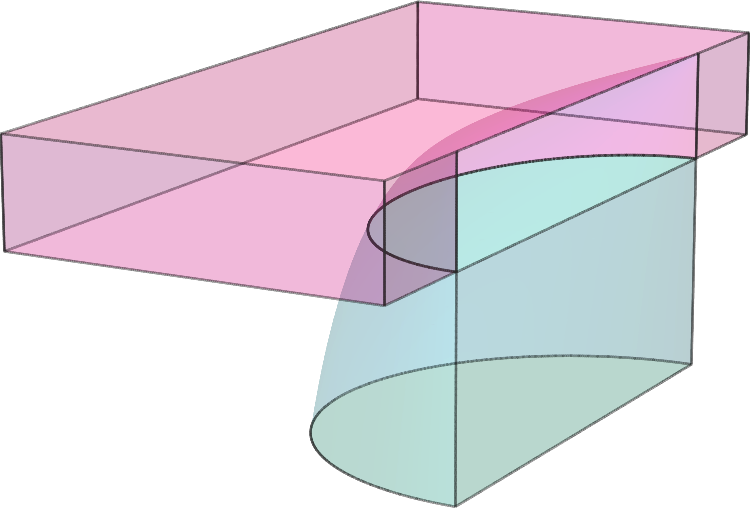}
        \caption{$c_1<0$ and $c_2>0$}
        \label{fig:a}
    \end{subfigure}
    \hfill
    \begin{subfigure}{0.49\textwidth}
        \includegraphics[width=\textwidth]{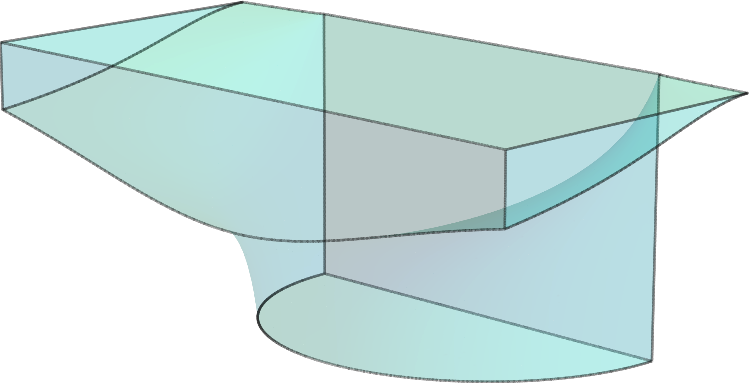}
        \caption{$c_1>0$ and $c_2<0$}
        \label{fig:b}
    \end{subfigure}
    \mbox{}\\[2ex]
    \begin{subfigure}{0.49\textwidth}
        \centering
        \includegraphics[width=.5\textwidth]{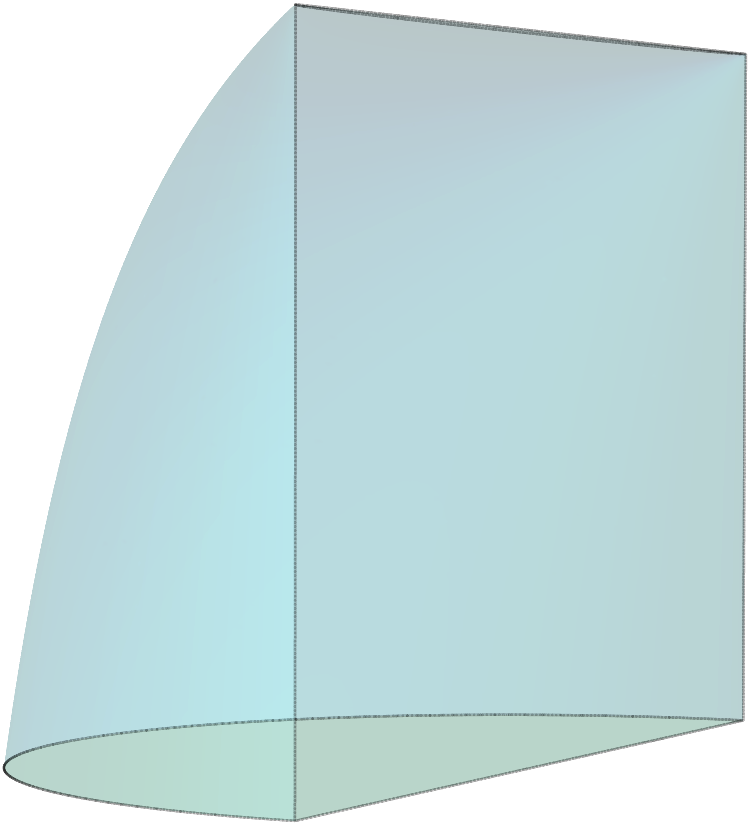}
        \caption{$c_1>0$ and $c_2>0$}
        \label{fig:c}
    \end{subfigure}
    \hfill
    \begin{subfigure}{0.49\textwidth}
        \includegraphics[width=\textwidth]{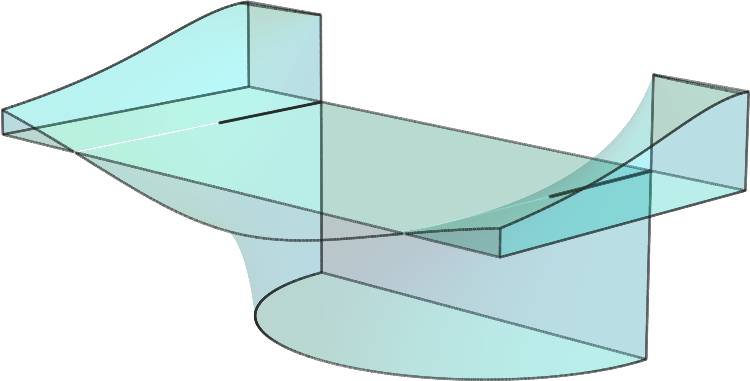}
        \caption{$c_1<0$ and $c_2<0$}
        \label{fig:d}
    \end{subfigure}
    \caption{Physical phase space for hybrid coupling according to different signs of $c_1$ and $c_2$.}
    \label{fig:hybridcoupling}
\end{figure}

So far, we have only discussed the cases in which either of the coupling constants is set to zero. Let us briefly turn our attention to the more complex example where both the coupling constant of the matter energy density, $\rho$, and that of the scalar field, $\phi$, are non-zero. The physical phase space according to the possible combinations of different signs of $c_1$ and $c_2$ is shown in Fig.~\ref{fig:hybridcoupling}. In particular, the phase space is unbounded in all instances except when $c_1>0$ and $c_2>0$, given in Fig.~\ref{fig:c}. This adds a layer of complexity since one would need to investigate the existence of potential critical points at infinity. Moreover, the dynamics of critical points present in these cases has been discussed in the two previous sections, hence, only the points at infinity would be of potential interest. 

The most interesting of these cases is when $c_1<0$ and $c_2>0$ because both points $D$ and $E$, given in Tab.~\ref{tab:generalfixedpoints}, exist. We note that, as one may expect, the physical phase space is a superposition of those in Fig.~\ref{fig:bound2} and Fig.~\ref{fig:bound3}, hence giving rise to Fig.~\ref{fig:a}. We will not provide a complete analysis of such models here, but note that this could lead to interesting results: for instance, early-time inflationary epochs along with late-time phantom dark energy transitioning to de Sitter expansion.
However, dealing with the unbounded phase space will require the introduction of new variables to study the points at infinity, which may be of physical interest. 

\clearpage

\section{Discussion}
\label{sec:disc}

The study of cosmological models through dynamical systems techniques has provided many interesting insights into the background dynamics of the Universe, particularly regarding the modelling and understanding of dark energy and dark matter interactions. This paper introduced new types of couplings involving a perfect fluid, building on Brown's variational principle for relativistic fluids, a canonical scalar field, and a geometrical term $\ourG$ related to the Ricci scalar. Models of this type offer a rich dynamical structure compared to many other models studied in the past. The key ingredient of our work is the interaction function of the form $f(n,\phi)\ourG$ which provides a non-minimal coupling between the geometry and the matter. Such couplings have not been considered before, neither with the pseudoscalar $\ourG$ nor the analogous teleparallel scalars $T$ or $Q$, marking the novelty of our construction.

Similar to previous work, we assume that the scalar field has an exponential potential $V(\phi) = V_0 \exp(-\lambda\phi)$. This ensures that the dimensionality of the phase space remains tractable; power-law potentials, for example, increase the dimensionality further, see~\cite{Bahamonde:2017ize}. 
Our setup has the main advantage that we work in a bounded three-dimensional phase space, which avoids the need to study points at infinity, thus simplifying our analysis. The adoption of standard variables also facilitates the comparison of our results with previously studied models.

The newly identified points $D$ and $E$ of Tab.~\ref{tab:stabilitygeneral} are of particular interest since the effective equation of state parameter is $-1$, modelling an inflationary phase or a saddle point with accelerated expansion, respectively. Even more interestingly, the fixed point $E$ was shown to lead to transient phantom behaviour, a relatively rare feature for models containing canonical scalar fields.

Our model naturally explains two accelerated epochs, early inflation and late-time accelerated expansion or dark energy. It also gives rise to additional phenomenologically interesting behaviour, such as dark energy with phantom crossing. A combination of all of these properties is rare to find in a unified framework, which motivates further investigations. Moreover, one can connect our results to other geometrical settings like teleparallel gravity or symmetric teleparallel gravity. This is achieved through the pseudoscalar $\ourG$ which naturally links to the teleparallel scalar or the non-metricity scalar, see~\cite{Boehmer:2023fyl,Jensko:2024bee}.  

In this work, we have restricted our attention to the matter-dominated case. However, the dynamical systems and stability analysis presented in Section~\ref{sec:cosmology} is valid for all $w$ and easily generalised to the case of multiple fluids.
Consequently, it would be interesting to study such models both radiation and matter. Regarding the specific choice of coupling function, this was  largely motivated by our desire to formulate a theory that could be studied explicitly. For more complicated coupling functions, this may no longer be the case. This follows essentially from the Friedmann constraint~\eqref{fried1}, which cannot be solved to eliminate either $y$ or $\sigma$ for complicated models. An alternative approach would be to instead eliminate the variable $x$. This would lead to a markedly different presentation than the dynamical systems studied in this work, but would connect more closely to the earlier works such as~\cite{Copeland:1997et}. In that case, one would expect difficulties in finding all the critical points as the new equations would depend on the coupling functions. Nonetheless, work in this direction appears to be promising.

A significant next step is to confront observational data to assess the viability of these models. It is clear that from a qualitative perspective, all of the models and couplings introduced in this work are of phenomenological interest and have the potential to fit astronomical data.
At the background level, supernovae and baryon acoustic oscillation data can then be used to probe the expansion history and derive constraints on our free parameters $(c_1,c_2)$ as well as predictions for key quantities such as the Hubble constant $H_0$. Since all the models considered in this paper affect the evolution of $H(z)$, we expect such constraints to have strong implications for our models. Moreover, in light of the recent DESI results~\cite{DESI:2025zpo,DESI:2025zgx}, which seems to favour phantom forms of dark energy, the scalar-field coupled models (which allow for phantom crossing) provide an interesting direction for further investigation. Finally, the study of cosmological perturbations would allow us to compare our model with a richer spectrum of observational data, such as Planck CMB data~\cite{Planck:2018vyg}. Given that our formulation is fully covariant and consistently derived from action principles, cosmological perturbations follow unambiguously from our approach. Such studies at the level of linear perturbations have been fruitful in other non-minimally coupled models using the Brown variational approach~\cite{Koivisto:2015qua}, but for our models extra care must be taken with the non-covariant terms $\ourG$ and $\ourB$.
A detailed study of these topics is planned for future works.

Our paper shows the potential of novel non-minimal couplings derived from a well-defined variational principle. The interaction between the scalar field, the perfect fluid, and the geometric term leads to interesting dynamical properties which range from scaling solutions to epochs of accelerated expansion at different times. 

\subsection*{Acknowledgements}   

Antonio d'Alfonso del Sordo is supported by the Engineering and Physical Sciences Research Council EP/R513143/1 \& EP/T517793/1. Erik Jensko is supported by the Engineering and Physical Sciences Research Council [EP/W524335/1]. 

\appendix

\section{Diffeomorphism invariance and conservation laws} \label{append_diff}
The covariant conservation of a generalised energy-momentum tensor is equivalent to the invariance of its action under infinitesimal diffeomorphisms. Given the Lagrangian  $f(n,s,\phi,\ourG,\ourB)$, where $\ourG$ and $\ourB$ are defined in Eqs.~(\ref{bulk})--(\ref{boundary}), performing an infinitesimal coordinate transformation $x^{\mu} \rightarrow x^{\mu} + \xi^{\mu}(x)$, leads to a change $\delta_{\xi}$ in the action given by~\cite{Boehmer:2021aji}
\begin{align}
    \delta_{\xi} S_{\textrm{int}} &= -\int \delta_{\xi} \left( \sqrt{-g} f(n,s,\phi,\ourG,\ourB) \right) \dd^4 x \nonumber \\
    &= -\int \left[ \mathcal{L}_{\xi}{\sqrt{-g}} f + \sqrt{-g} \left( \pdv{f}{n} \mathcal{L}_{\xi} n + \pdv{f}{s} \mathcal{L}_{\xi} s + \pdv{f}{\phi} \mathcal{L}_{\xi} \phi
 +\pdv{f}{\ourG} \mathcal{L}_{\xi} \ourG  + \pdv{f}{\ourB} \mathcal{L}_{\xi} \ourB \right) \right] \dd^4x \, , \nonumber \\    
  &= -\int \left[ \partial_{\mu}\left( \sqrt{-g} f  \xi^{\mu} \right)  + \sqrt{-g} \Big(\pdv{f}{\ourG}-\pdv{f}{\ourB}\Big) E_{\lambda}{}^{\mu \nu} \partial_{\mu} \partial_{\nu} \xi^{\lambda} \right] \dd^4x \, , \nonumber \\ 
    &= \textrm{`boundary terms'} + \int \partial_{\mu} \partial_{\nu} \left( \sqrt{-g} E_{\lambda}{}^{\mu \nu} \Big(\pdv{f}{\ourB} -\pdv{f}{\ourG} \Big) \right) \xi^{\lambda} \dd^4x \, . \label{diff} 
\end{align}
In the second line, we have applied the Lie derivative for the infinitesimal diffeomorphism $\delta_{\xi} = \mathcal{L}_{\xi}$ with the standard expressions for scalars and the metric determinant. In the subsequent line, we have also made use of the formula for the Lie derivative of the bulk and boundary terms~\cite{Boehmer:2021aji,Jensko:2023lmn} 
\begin{equation}
\mathcal{L}_{\xi}\ourB = \xi^{\lambda} \partial_{\lambda} \ourB - E_{\lambda}{}^{\mu \nu} \partial_{\mu} \partial_{\nu}\xi^{\lambda} \, , \qquad \mathcal{L}_{\xi}\ourG = \xi^{\lambda} \partial_{\lambda} \ourG + E_{\lambda}{}^{\mu \nu} \partial_{\mu} \partial_{\nu}\xi^{\lambda} \, ,
\end{equation}
where the rank-three object $E^{\lambda \mu \nu}$ is defined in~(\ref{E}).
In transitioning to the final line we have integrated by parts twice, with all boundary terms implicitly included in the first term vanishing for appropriate boundary conditions.
The additional terms in the total action are the usual Einstein-Hilbert term and the scalar field action, both of which are diffeomorphism invariant scalars by construction. It therefore follows that if the remaining integrand of~(\ref{diff}) vanishes, then $\delta_{\xi}S_{\textrm{tot}}=0$. Employing Noether's theorem then yields the generalized conservation law given in equation~(\ref{noncons}). This is seen explicitly by letting the metric variation be generated by an infinitesimal diffeomorphism $\delta_{\xi} g_{\mu \nu} = \mathcal{L}_{\xi} g_{\mu \nu}$ along with the definition of $ \mathcal{L}_{\xi} g_{\mu \nu}$. Finally, we note that the gravitational part gives rise to the Einstein tensor,  which is divergence-free due to the contracted Bianchi $\nabla^{\mu}G_{\mu \nu} =0$, and the resulting terms are precisely~(\ref{noncons}).

We therefore see that the conservation law~(\ref{noncons}) is equivalent to the coordinate-dependent condition
\begin{equation} \label{noncons2}
    \partial_{\mu} \partial_{\nu} \Big( \sqrt{-g}E_{\lambda}{}^{\mu \nu}{} (f_{,\ourB} - f_{,\ourG})\Big) = 0 \, .
\end{equation}
The diffeomorphism invariant limit of these models is $f(R,n,s,\phi)$ gravity and the above constraint vanishes identically.
On cosmological backgrounds in Cartesian coordinates\footnote{Note that other choices of cosmological coordinates would not lead to the constraint~(\ref{noncons2}) vanishing identically~\cite{Jensko:2024bee}.}, the bulk and boundary terms take the form
\begin{equation}
    \ourG = -\frac{6H^2}{N^2} \, , \qquad \ourB = \frac{6(3H^2 N + N \dot{H} - H \dot{N} )}{N^3} \, , \label{appendB}
\end{equation}
the non-vanishing components of $E^{\lambda \mu \nu}$ are
\begin{equation}
    E_{0}{}^{ii} = 2 \frac{H}{a^2} \, , \qquad E_{i}{}^{i0} = -4 \frac{H}{N} \, , \qquad E_{i}{}^{0i} = \frac{4HN + \dot{N}}{N^2} \, ,
\end{equation}
and the scalars $n$, $s$ and $\phi$ are functions of time. It then directly follows that for these values the constraint~(\ref{noncons2}) vanish, independent of the choice of $f$. We are therefore free to study any choice of model without additional constraints in these cosmological coordinates. Moreover, clearly $f(n,s,\phi,\ourG)$ is a subset of $f(n,s,\phi,\ourG,\ourB)$ and so the analysis above applies to the models studied in this work. This can also be verified directly from the metric field equations, where Eq.~(\ref{noncons}) imposes no additional constraints on the choice of model $f$ on these cosmological backgrounds. 

\section{Singularities and limiting behaviour}
\label{app:sing}

The dynamical system defined by the equations~(\ref{eq:gen1})--(\ref{eq:gen3}) generically exhibits singular behaviour in the limit $z \to 1$, where the Hubble functions diverges $H \rightarrow \infty$, see~(\ref{eq:zlim}). Moreover, physical quantities such as the deceleration parameter~(\ref{eq:q}) also display singular behaviour as $z \to 1$. 

Examining this more closely, let us start with~(\ref{eq:gen1})--(\ref{eq:gen3}) and make a perturbative expansion around $z=1$ by setting $z=1-\epsilon$, $\epsilon \ll 1$, to ensure we approach $z=1$ from below. This expansion yields
\begin{align}
    x' &= \frac{c_2}{2\epsilon} \frac{c_2 \sqrt{6} y^2 \lambda + c_1 ( \sqrt{6} \lambda(x^2 -y^2) - 3 x (1+w))}
    {c_1(x^2+y^2) - c_2 y^2}y^2  +  
    \mathcal{O}(\epsilon^0) \,, 
    \label{eq:append1}\\
    y' &= -\frac{c_1 c_2 }{2\epsilon} \frac{3+3w -2 \sqrt{6} x \lambda}{c_1(x^2+y^2) - c_2 y^2} y^3 + 
    \mathcal{O}(\epsilon^0) \,, 
    \label{eq:append2}\\
    z' &= \mathcal{O}(\epsilon^0) \,. 
    \label{eq:append3}
\end{align}
In order for the $1/\epsilon$ terms to vanish for all $x$ and $y$, one must require $c_2 = 0$. We therefore note that $c_2=0$ gives the only non-trivial, regular class of models for $z \to 1$. On the other hand, if $y \to 0$ sufficiently fast as $z \to 1$, one would also be able to discuss a meaningful model. Let us perform the same expansion on the declaration parameter~(\ref{eq:q}), which yields
\begin{equation}
    q = -\frac{1}{2\epsilon} \frac{c_1 c_2 y^2 \big(3+3w - 2 \sqrt{6} x \lambda \big)}{c_1 (x^2+y^2) - c_2  y^2}+ 
    \mathcal{O}(\epsilon^0) \, .
\end{equation}
For this to be regular when $\epsilon \rightarrow 0$ for all $x$ and $y$, one now obtains either $c_1=0$ or $c_2=0$. 

\begin{figure}[!htb]
    \centering
    \includegraphics[width=0.40\textwidth]{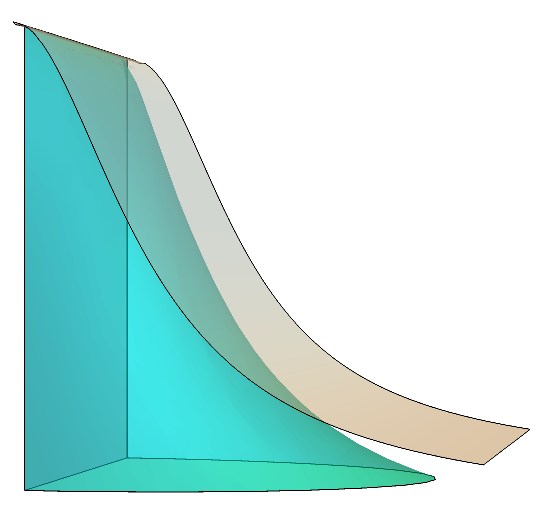}
    \caption{Physical phase (cyan) of the $c_1=0$, $c_2>0$ model including the surface $S$ (gray) where the system is singular. All axes and labels are suppressed for simplicity.}
    \label{fig:sheet}
\end{figure}

Focussing now on the scalar-field coupling model $c_1=0$ and $c_2 > 0$ given in Section~\ref{sec:scalarfield}, it follows that the the dynamical system~(\ref{eq:gen1})--(\ref{eq:gen3}) is not well-defined for the limit $z \to 1$. Explicitly, the $x'$ equation diverges, as can be seen from Eq.~(\ref{eq:append1}). The equations of the system are singular when $1 - c_2 y^2 z/(1-z) = 0$. Thus, we define the surface $S$ by 
\begin{align}
    S(x,y,z)=1 - c_2 y^2 \frac{z}{1-z} = 0 \,,
    \label{singularsurface}
\end{align}    
which is a 2D sheet. This sheets lies `above' the physical phase space. In the limit that $z \to 1$, we have $y \to 0$. This means that this sheet is attached to the phase space along the line $z=1$, $y=0$, as illustrated in Fig.~\ref{fig:sheet}. The singular line $z=1$, $y=0$ is therefore of particular importance for this model.

Taking the limit $z \rightarrow 1$ along trajectories with $y=0$ (which is always within the physical phase space) yields consistent results, as can also be seen from the perturbative expansions~(\ref{eq:append1})--(\ref{eq:append3}), which remain regular for $y=0$. This can be justified by assuming that $y \to 0$ faster than $z \to 0$, such that trajectories remain below $S$ at all times. The singular points and stability analysis obtained by taking these limits (with $y \to 0$ and $z \to 0$) were given in Tab.~\ref{tab:limits}. The numerical simulations and evolution plots also confirm that this limit is valid for physically realistic trajectories in the phase space.

\addcontentsline{toc}{section}{References}
\bibliographystyle{jhepmodstyle}
\bibliography{bib_new-inspireized}

\end{document}